\documentclass[ip]{jpsj3}
\usepackage{txfonts}
\usepackage{color}
\usepackage{bm}
\usepackage{url}
\usepackage{overcite}

\title{Drawing Phase Diagrams of Random Quantum Systems by Deep Learning the Wave Functions}

\author{Tomi Ohtsuki\thanks{ohtsuki@sophia.ac.jp} and Tomohiro Mano }
\inst{
Physics Division, Sophia University, Chiyoda-ku, Tokyo 102-8554, Japan} %

\abst{
Applications of neural networks to condensed matter physics are becoming popular and beginning to be well accepted.
Obtaining and representing the ground and excited state wave functions are examples of such applications.
Another application is analyzing the wave functions and determining their quantum phases.
Here, we review the recent progress of using the multilayer convolutional neural network, so-called deep learning, to determine the quantum phases
in random electron systems.  After training the neural network by the supervised learning of wave functions in restricted parameter regions in known phases, the neural networks can
determine the phases of the wave functions in wide parameter regions in unknown phases;
hence, the phase diagrams are obtained.  We demonstrate the validity and generality
of this method by drawing the phase diagrams of
two- and higher dimensional Anderson metal-insulator transitions and
quantum percolations as well as disordered topological systems such as three-dimensional
topological insulators and Weyl semimetals.
Both real-space and Fourier space wave functions are analyzed. 
The advantages and disadvantages over conventional methods are discussed.}

\begin{document}
\maketitle


\section{Introduction}
More than seven decades have passed since McCulloch and Pitts studied artificial neurons\cite{McCulloch43}.
Using artificial neurons, Rosenblatt proposed a layered neural network called the perceptron,\cite{Rosenblatt58}
which consists of input and output layers together with intermediate layers.
The connections between layers, called weight parameters, can be changed to reproduce the correct
input--output relations.  This process of tuning the parameters is essentially ``machine learning."
The idea of the multilayer perceptron\cite{Fukushima80} is simple and can be applied to solve
many problems,
but it is only in the last decade that our computers have become powerful enough to solve complicated
problems via large-scale multilayer neural networks, called deep learning.\cite{LeCun15,Mnih15,Silver16,Goodfellow16}

Computational physics has successfully solved many problems in solid-state physics, and it is
natural to use machine learning, including deep learning, to tackle complicated problems.
In fact, in the last few years, there has been growing interest in applying machine learning to condensed matter physics.\cite{Carleo19}
Machine learning is used  to obtain and represent ground- and excited-states wave functions,
\cite{Carleo16,Saito17,Nomura17,Saito17b,Saito18,Gao17,Deng17,Cai18,Carleo18,Choo18,Luo18,Amin18,Liang18,Yao19,Shi19,Wu19,Jia19,Ferrari19,He19,Irikura19,Moreno19}
even the states in open systems\cite{Yoshioka19, Vicentini19,Nagy19},
to improve Monte Carlo simulations,\cite{Huang17,Nagai17,Shen18,Nagai18,Yoshioka18a,Li19b,Yang19b,Nagai19,Zhao19,Xu19,Theveniaut19,Nicoli19,Schwarz17}
to construct effective potential/Hamiltonian,\cite{Takahashi17,Wenwen17,Fujita18,Mills18,Arai18,Sidky18,Bartok18,Vlcek18,Hashimoto18,Nagai19a,Jinno18,Hu19,Ceriotti18,Liu19b,Caro19,Babaei19,Mannodi-Kanakkithodi19,Rosenbrock19,Zhang19b,Qian19,Thomas19,Berressem19,Wang19a,Schmidt19,Byggmastar19,Nakamura19a,Freitas19,Zeni19,Sauceda19,Patra19,Sivaraman19}
and to accelerate density functional calculation,\cite{Li16,Suwa19}
as well as to analyze experimental data such as
 X-ray patterns \cite{Xu18,Vecsei18,Utimula18,Li19a,Greco19},
photoemission,\cite{Drera19}  microscope images\cite{Ziatdinov17,Maksov18,Picard19,Zhang19c,Ede19,Masubuchi19,Zheng19,Burzawa19} and snapshots of
momentum-space density images of cold atom systems.\cite{Rem18}
It can be used to efficiently predict material properties such as the energy landscape,\cite{Olsthoorn18,Teichert19,Noe18,Deringer18} crystal symmetry,\cite{Liu19a} lowest energy level,\cite{Pilati19} magnetic properties,\cite{Pham17,Dam18,Pham18,Exl18,Nakamura19,Nelson19,Samarakoon19,Rzadkowski19}
and critical temperatures\cite{Stanev18,Ramprasad19,Konno18},
which can be used for material design
\cite{Kiyohara16,Oliynyk16,Lee16,Mills17,Ramprasad17,Seko18,Gubaev18,Pilozzi18,Xie18,Ye18,Bartok17,Dai18,Iwasaki18,Iwasaki19,Yang18,Hanakata18,Sagiyama19,Han19,Boattini19,Wakabayashi19,Ashida19,Kiyohara19,Sun19,Oda19,Brunton19,Barker19,Zhang19c,Goodall19,Xu19a,Lu19,Jinnouchi19,Haug19} in material informatics.\cite{Takabe14,Seko15,Terakura17,Tsujimoto18,Matsumoto18,Gao18,Park19,Yamaji19,Schleder19,Nguyen19a,Shao19,Claussen19,Rigo19}
The method is also used for predicting light propagation in photonic crystals without
solving the Maxwell equations,\cite{Tahersima18,Asano18}
which will lead to efficient photonic device design\cite{Wiecha19,Li19c}.

One of the  important applications of machine learning is to
extract features from data, and  in fact,
unsupervised trainings of spin \cite{Wang16,Wetzel17a,Hu17,Wang17,Zhao18,Kiwata19,Durr19,Alexandrou19,Woloshyn19,Greplova19,Shirinyan19}
and hard sphere systems\cite{Jadrich18b,Jadrich18a} have identified different phases of matter.\cite{Vargas-Hernandez18,Xu19b}
The features of states, however, are often very complicated in the case of disordered and correlated
electron systems.
Applications of  neural networks obtained by supervised training, which has proved to be very powerful
in image recognition, are therefore expected to work well for this purpose.
Thermal phase transitions of spin systems such as Ising and XY models as
well as quantum phase transitions of frustrated spin systems have been
detected successfully
\cite{Carrasquilla17,Carrasquilla17b,Tanaka17,Nieuwenburg17,RodriguezNieva18,Liu18b,Beach18,Liu18c,RichterLaskowska18,Giannetti18,Li18d,Greitemann19,Holanda19,Aoki19,Choo19,Munoz-Bauza19,Canabarro19,Kiwata19,Decelle19,Wu19a,Zhang19e}.
Critical exponents of the Ising model\cite{Carrasquilla17,Carrasquilla17b,KochJanusz18,Efthymiou18} as well as the classical
percolation transition\cite{Zhang18b,Ni19} are also estimated by machine learning.
Novel magnetic configurations such as the skyrmion are recognized using neural networks.\cite{Iakovlev18,Singh19}
The thermal magnetic phase transition of interacting electron systems such as three-dimensional (3D) Hubbard models
 is analyzed using neural networks\cite{Chng17,Chng18},
and the quantum phase transition of the Hubbard model is detected via the convolutional neural network (CNN).\cite{Broecker17,Broecker17b,Costa17,Dong19,Ma18,Shinjo19}
Quantum loop topography\cite{Zhang17a} is used for studying 
the quantum phase transition of spin liquids as well as the
superconductivity in the Hubbard model.\cite{Zhang17b,Zhang19}

Quantum phase transitions in random electron systems are also interesting to study.
Random electron systems show the Anderson-type metal--insulator transition,\cite{Anderson58,Kramer93,Evers08} (so-called Anderson transition, also called
delocalization--localization transition), quantum percolation transition,\cite{StaufferBook}
 topological--nontopologogical transitions,\cite{shindou09prb,Kobayashi13}
 and semimetal--insulator\cite{Luo18a,Luo18b}
and semimetal--metal transitions.\cite{Fradkin86,Goswami11,Kobayashi14,Liu16}
The wave functions of the random/interacting quantum system are difficult to analyze owing
 to the large fluctuations of the wave functions, but a
trained CNN has been shown to detect quantum phase transitions.
It can detect topological states in one-dimensional (1D) systems,\cite{Balabanov19,Tsai19}
two-dimensional (2D) Anderson transition and topological transitions such
as the band--to--Chern insulator transition\cite{Tomoki16,Zhang17a,Carvalho18,Cheng18,Caio19},
2D topological superconductor transition\cite{Yoshioka18},
higher order topological insulators\cite{Araki19},
3D Anderson and quantum percolation transitions\cite{Mano17},
as well as 3D topological phase transitions\cite{Tomoki17,Bednik19,Lian19,Mano19} such as topological insulators and Weyl semimetals.
Quantum chaos\cite{haakebook} is related to a random electron system, which is also studied using
 neural networks.\cite{Kharkov19,Ma19}

The interplay of randomness and interaction is attracting renewed interest from the view point of
many-body localization,\cite{BASKO20061126,Gornyi05,Nandkishore15,ALET2018498,Abanin19} where the hypothesis of ``eigenstate thermalization" no longer applies,
and the machine learning is again shown to be powerful in recognizing whether the phase thermalizes.\cite{Schindler17,Berkovits18,Venderley18,Huembeli18,Hsu18,Zhang18c,Doggen18,Gray18,Theveniaut19a,Rao18a,Matty19,Hartmann19,Nieuwenburg19}

In this paper, we review the application of the CNN to draw phase diagrams in random quantum systems.
In the next section, we explain the methods, followed by a section on models and results,
where the Anderson metal--insulator transitions and quantum percolation transitions in various dimensions,
as well as the 3D topological insulator and  Weyl semimetal transitions, are discussed.
The last section is devoted to summary and concluding remarks.

Providing an exhaustive overview of the extensive literature on machine learning would be an exacting task,
so here we focus on drawing the phase diagrams of quantum phase transitions in random systems.
We do not pretend to give an exhaustive overview, and apologize that many aspects of the machine
learning approaches in condensed matter physics will not be covered.


\section{Methods}
\label{sec:method}
\begin{figure}[htb]
  \begin{center}
\includegraphics[angle=0,width=1.0\textwidth]{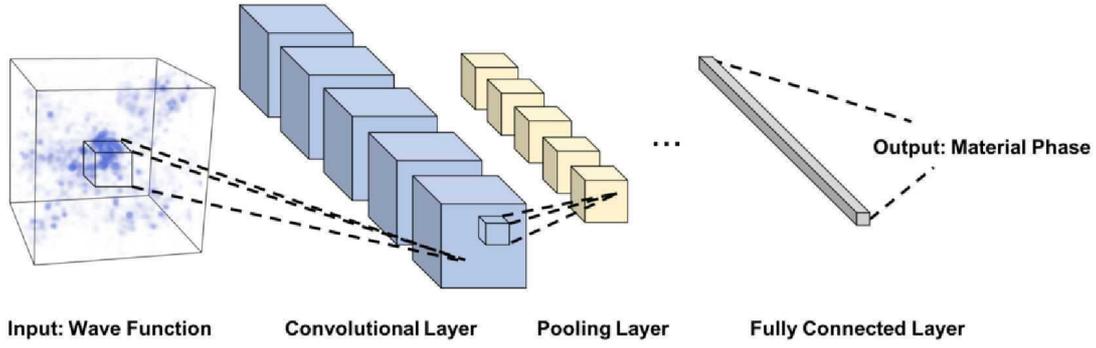}
 \caption{(Color online) Schematic of the CNN structure given 3D input.
 Convolutional layers and pooling layers are repeated, and finally, the material phase
to which the eigenfunction belongs is output through a fully connected layer.
Here, the channel number of the first layer $C_1$ is 5.
Cases of the $d$-dimensional input, 2D, 
four-dimensional (4D) and higher, are realized with a similar configuration.  }
\label{fig:CNNSchematic}
\end{center}
\end{figure}

To draw the phase diagrams, we use the CNN consisting of three types of layers: convolutional layers, pooling layers, and fully connected layers.
The basic structure of the CNN is illustrated in Fig.~\ref{fig:CNNSchematic}.
This type of CNN has proved to be very powerful for image recognition.  A famous example is LeNet.\cite{LeCun15}
Given the input to the first layer, the output of one layer propagates to the input of the next layer, and finally, the output of the last layer is obtained.
The CNN is therefore a type of feedforward network.
The detailed process of the CNN is as follows.

We consider the electron density $|\psi(\bm{x}_i)|^2$ at $\bm{x}_i$ (site index $i$) as input $u_{i}^{(0)}$.
In the first convolutional layer, cells with a certain size are cut out from the input $\bm{u}^{(0)}$
(see the small cube inside ``Input: Wave Function" in Fig.~\ref{fig:CNNSchematic})
 and transformed by
\begin{equation}
\label{eq:conv}
u_{j,k}^{(1){\prime}}=\bm{W}_{k}^{(1)} \cdot  \bm{u}_{j}^{(0)}+b_k^{(1)}\,,
\end{equation}
where $\bm{u}_{j}^{(0)}$ denotes the component of $u_{i}^{(0)}$ in the $j$th cell, which is originally a  tensor of rank-$d$ ($d$ being the dimensionality of the system) but arranged one-dimensionally, and $\bm{W}_{k}^{(1)}$ and $b_k^{(1)}$
are the weight parameter of channel $k (1\le k\le C_1)$ and the bias parameter for channel $k$, respectively.
The weight parameter $\bm{W}_{k}^{(1)}$ has the same dimension as $\bm{u}_{j}^{(0)}$ and does not depend on the position $j$ at which the cell is cut out.
During the training of the CNN,   $\bm{W}$ and $\bm{b}$ are optimized to reproduce the input (eigenfunction)--output (material phase) relations.
The convolution process corresponds to extracting the local features of the input data.
We stride the position to cut out the cell and obtain the output $\bm{u}^{(1){\prime}}$ so that we obtain $C_1$ images from an input image.
We then apply the rectified linear unit (ReLU) to the output,
\begin{equation}
u_{j,k}^{(1)}=\text{max}(0,u_{j,k}^{(1){\prime}})\,,
\end{equation}
to obtain $u_{j,k}^{(1)}$, where max$(0,x)$ acts as an activation function expressing the firing of neurons.

Note that the size of the cell, the stride value, and the numbers of channels are hyperparameters that
 cannot be optimized by training,
and need to be chosen appropriately a priori.
In general, the selection of hyperparameters has an effect on learning accuracy in the training.

In the second and subsequent convolutional layers, the convolution process is performed over all channels.
Then the transformation from the $(n-1)$th layer to the $n$th layer with channel $k \,(1\le k\le C_n)$ is described as
\begin{equation}
\label{eq:conv2}
u_{j,k}^{(n){\prime}}=\sum_{m=1}^{C_{n-1}}
\bm{W}_{m,k}^{(n)} \cdot \bm{u}_{j,m}^{(n-1)}+b_k^{(n)}\,,
\end{equation}
where $C_n$ and $C_{n-1}$ denote the total number of channels in the $n$th and $(n-1)$th layers,
respectively.
As with the first convolutional layer, we apply the ReLU to the output as an activation function,
$u_{j,k}^{(n)}=\text{max}(0,u_{j,k}^{(n){\prime}})$.

In the pooling layer, located mainly after the convolutional layer, the maximum value in the cell is chosen,
\begin{equation}
u_{j,k}^{(n+1)}=\text{max} (u_{i,k}^{(n)}\:|\:i \in j\,\rm{th\;cell}) .
\end{equation}
The number of channels is the same before and after the layer, since the sum over channels is not taken.
The pooling process corresponds to removing noise and is useful for reducing the dimension of the input data.

In the fully connected layer, located mainly before the final output of the CNN,
the multi-dimensional vector output from the convolutional layer or the pooling layer is flattened to the
1D vector $\bm{u}$.
Then, it is transformed by
\begin{eqnarray}
u_{q}'&=&\text{max} \left(0,\bm{W}_{q} \cdot \bm{u}+ b_q \right)\,, \\
u_{r}''&=&\sum_{q}
W_{q,r} u_{q}'+b_r'\,,
\label{eq:finalPhase}
\end{eqnarray}
where $q$ denotes the component of the vector $\bm{u}'$ and $r$ denotes the index of each material phase.
(We consider the fully connected layer consisting of two layers such as LeNet, but in a simple case, it is realized in one layer,
$u_{r}''=\bm{W}_{r} \cdot \bm{u} + b_r$.)
In the case of the Anderson model, $r=0$ and 1 correspond to the localized and delocalized phases, respectively.

In the final stage, we apply the softmax function to the last output $u_{r}''$,
\begin{equation}
u_{r}^{\text{(out)}}=\frac{\exp{u_{r}''}}{\sum_{r'}\exp{u_{r'}}''},
\end{equation}
and obtain the final output $u_{r}^{\text{(out)}}$, 
which represents the ``confidence" or ``probability" $P_r$  that the eigenfunction belongs to the phase of index $r$.

To obtain a meaningful final output $u_{r}^{\text{(out)}}$, it is necessary to optimize $\bm{W}$ and $\bm{b}$ in each layer.
In classification problems such as quantum phase determination,
it is appropriate to update these parameters to minimize the cross entropy,
\begin{equation}
S=-\sum_{r,i} P'_{r,i} \log{P_{r,i}},
\end{equation}
closely related to the maximum likelihood estimation\cite{Goodfellow16},
 where $P'_{r,i}$ is a desired output value paired with the input data $i$, i.e., the correct phase to which the eigenfunction $i$ belongs.
Various methods for efficiently minimizing $S$ have been proposed\cite{Goodfellow16} such as AdaGrad, RMSProp, AdaDelta, and Adam.
The gradient calculation required for these methods is performed by back
propagation\cite{Amari67,Rumelhart1986}, the application of the chain rule to the neural network,
to avoid massive numerical differentiations that require high computational costs.

To train the neural network, we have to prepare the correctly labeled eigenfunctions (training set), so that  $\bm{W}$ and $\bm{b}$ are optimized automatically, and the CNN captures the features of the eigenfunctions.
Once the CNN is trained, it is expected to determine the correct phase to which the unlabeled eigenfunctions belong.
To confirm the performance of the CNN,
we regard 10\% of the training set as the validation test set and train the CNN with the remaining 90\%.
We feed the input from the validation test set to the CNN and see whether it correctly 
reproduces the label of the input, i.e., the material phase.
In the following results, we have confirmed that the validation accuracy is over 97\%.
A high validation accuracy indicates that the CNN sufficiently captures the features of the eigenfunctions from the training set
and is expected to correctly judge unknown inputs as well.

For the numerical simulation, we consider the lattice model and diagonalize the Hamiltonian with various parameters.
Disorder in the Hamiltonian is generated from random numbers by the Mersenne Twister algorithm.\cite{Matsumoto98}
When a limited range of energy is needed, mainly for the training set,
we use the sparse matrix diagonalization algorithm Intel MKL/FEAST.\cite{Polizzi09}
When all the eigenergies and eigenvectors are needed, we use the standard linear algebra package LAPACK.\cite{lapack}
In most of the cases, we use real-space wave functions as the input, but in some cases,
we use Fourier-transformed wave functions
obtained through discrete Fourier transformation from real-space ones.

Depending on the phases, the wave functions show specific features.
Examples of wave functions in various material phases are shown in Sect.~\ref{sec:wf}.

We construct 2D and 3D CNNs using Keras\cite{chollet2015keras} as the frontend and TensorFlow\cite{tensorflow} as the backend,
whereas only TensorFlow is used for a 4D CNN.
To construct 4D convolutional and pooling layers, which are not prepared as a function in TensorFlow,
we simply repeat the 3D convolutional and pooling layers so that they are equivalent to 4D layers.
The detailed network hyperparameters of our CNN are shown in Sect.~\ref{sec:hyperparameters}.


\section{Models and Results}
\label{sec:results}
We train and use the CNN to analyze the eigenfunctions obtained by diagonalizing the
tight-binding Hamiltonian on a hypercubic lattice,
\begin{equation}
\label{eq:Hamiltonian}
H=\sum_{\bm{x}}
c_{\bm{x}}^\dagger  \, v_{\bm{x}} \,  c_{\bm{x}}-
\sum_{\langle \bm{x},\bm{x}'\rangle} 
c_{\bm{x}}^\dagger \, V_{\bm{x},\bm{x}'} \,
 c_{{\bm{x}}'}\,,
\end{equation}
where $\bm{x}$ indicates the position in a $d$-dimensional space,  $c_{\bm{x}}^\dagger $($c_{\bm{x}}$)
the creation (annihilation) operator at site $\bm{x}$, and
$V_{\bm{x},\bm{x}'}$ the transfer between sites $\bm{x}$ and $\bm{x}'$ with $\langle \bm{x},\bm{x}'\rangle$
restricting the transfer only between the nearest neighbors.
$v_{\bm{x}}$ is the random potential at site $\bm{x}$.
In the following, we consider square (2D), cubic (3D) and four dimensional (4D) hypercubic lattices.
When we include spin and orbital degrees of freedom, $c_{\bm{x}}$ becomes a vector and
$V_{\bm{x},\bm{x}'}$ a matrix.

The universality class of the random electron
system\cite{Wigner51,Dyson61,Dyson62,Efetov80,Hikami81,Altland97,Zirnbauer96} is determined by the 
basic symmetries of the Hamiltonian, such as time-reversal symmetry (TRS) and spin-rotation symmetry (SRS).
Systems with broken TRS belong to the unitary class.
Systems with both TRS and SRS belong to the orthogonal class, whereas those with TRS but broken SRS belong
to the symplectic class.
In our model, we change the universality class by modifying the transfer $V_{\bm{x},\bm{x}'}$.
In the absence of a magnetic field and spin--orbit interaction, we take $V_{\bm{x},\bm{x}'}=1$,
and the system belongs to the orthogonal class.
The choice $V_{\bm{x},\bm{x}'}=\exp(i\theta_{\bm{x},\bm{x}'})$ describes the presence of a magnetic field,
which breaks TRS; hence, the systems belong to the unitary class.
To discuss the effect of spin-orbit interaction, $V_{\bm{x},\bm{x}'}$ is set to 
SU(2) matrices\cite{Ando89,Evangelou87,Asada02,Asada04}  with the site potential $v_{\bm{x}}$ independent of spin.
In this case, TRS is preserved but SRS is broken; hence, the systems belong to the symplectic class.

Similar tight-binding models are used to discuss topological materials:
in the case of the 3D topological insulator, $V_{\bm{x},\bm{x}'}$ is set to be proportional to Dirac gamma matrices,\cite{Liu:3DTI}
whereas in the case of the 3D Weyl semimetal, $V_{\bm{x},\bm{x}'}$ is set to be proportional to Pauli matrices.\cite{Chen15,Liu16}  See Eqs. (\ref{eqn:H}) and (\ref{eq:WSMtb1}).

\begin{figure}[htb]
\centering
    \begin{tabular}{c}
 \hspace{-0.7cm}   
     \begin{minipage}{0.33\hsize}
  \begin{center}
   \includegraphics[width=34mm]{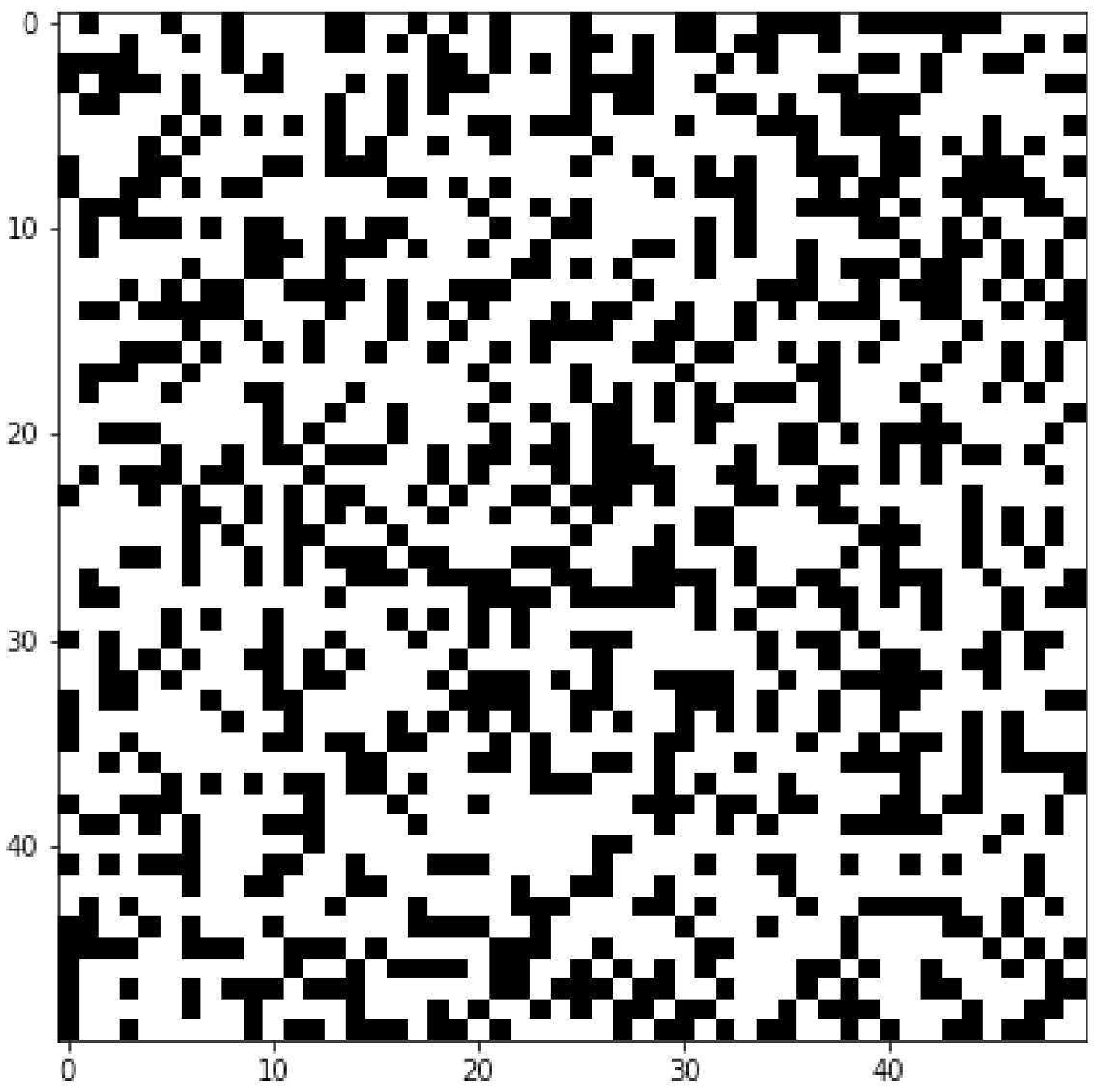}
   \hspace{1.6cm} (a)
     \end{center}
 \end{minipage}

     \begin{minipage}{0.33\hsize}
  \begin{center}
   \includegraphics[width=34mm]{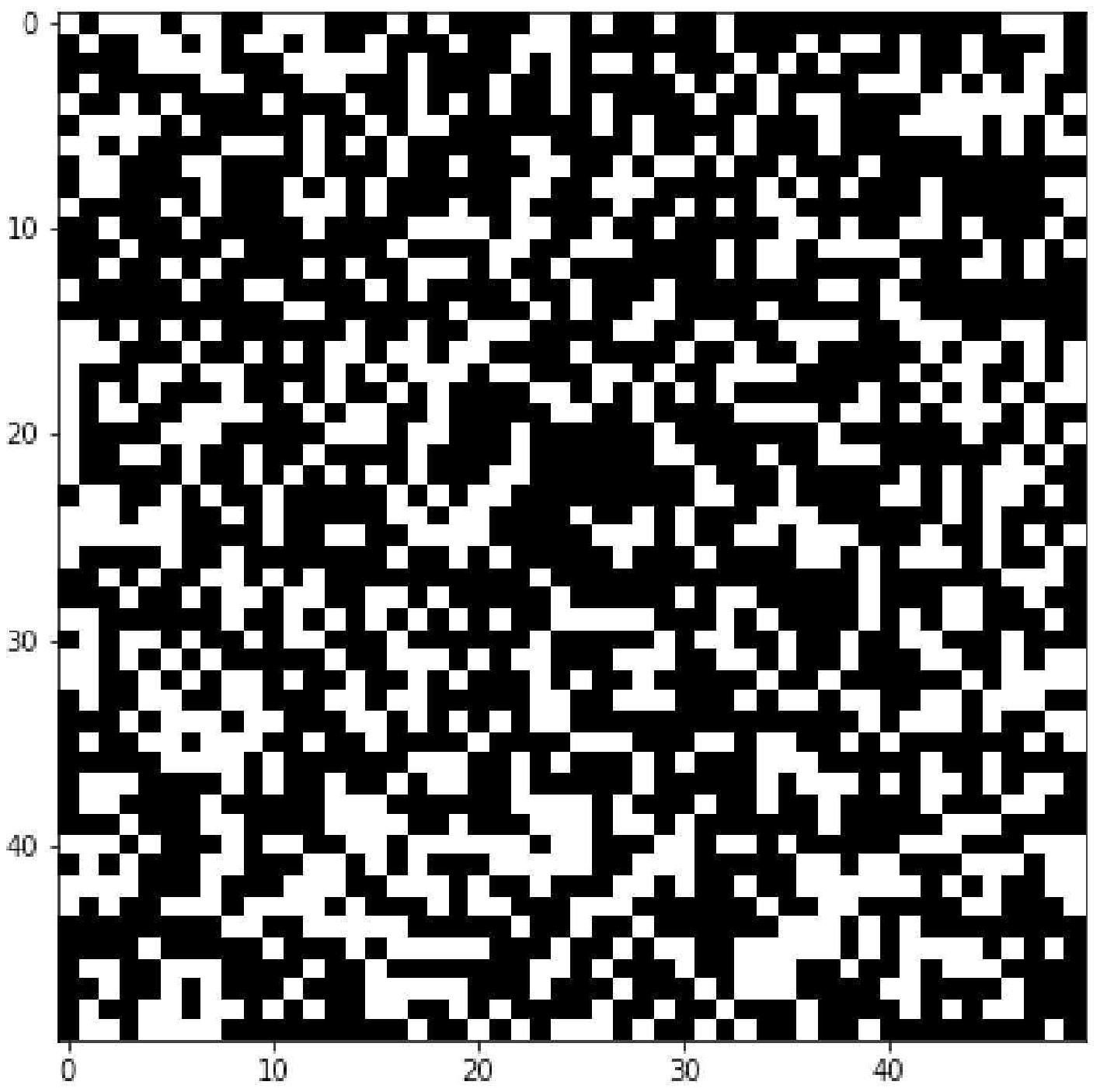}
   \hspace{1.6cm} (b)
     \end{center}
 \end{minipage}

     \begin{minipage}{0.33\hsize}
  \begin{center}
   \includegraphics[width=34mm]{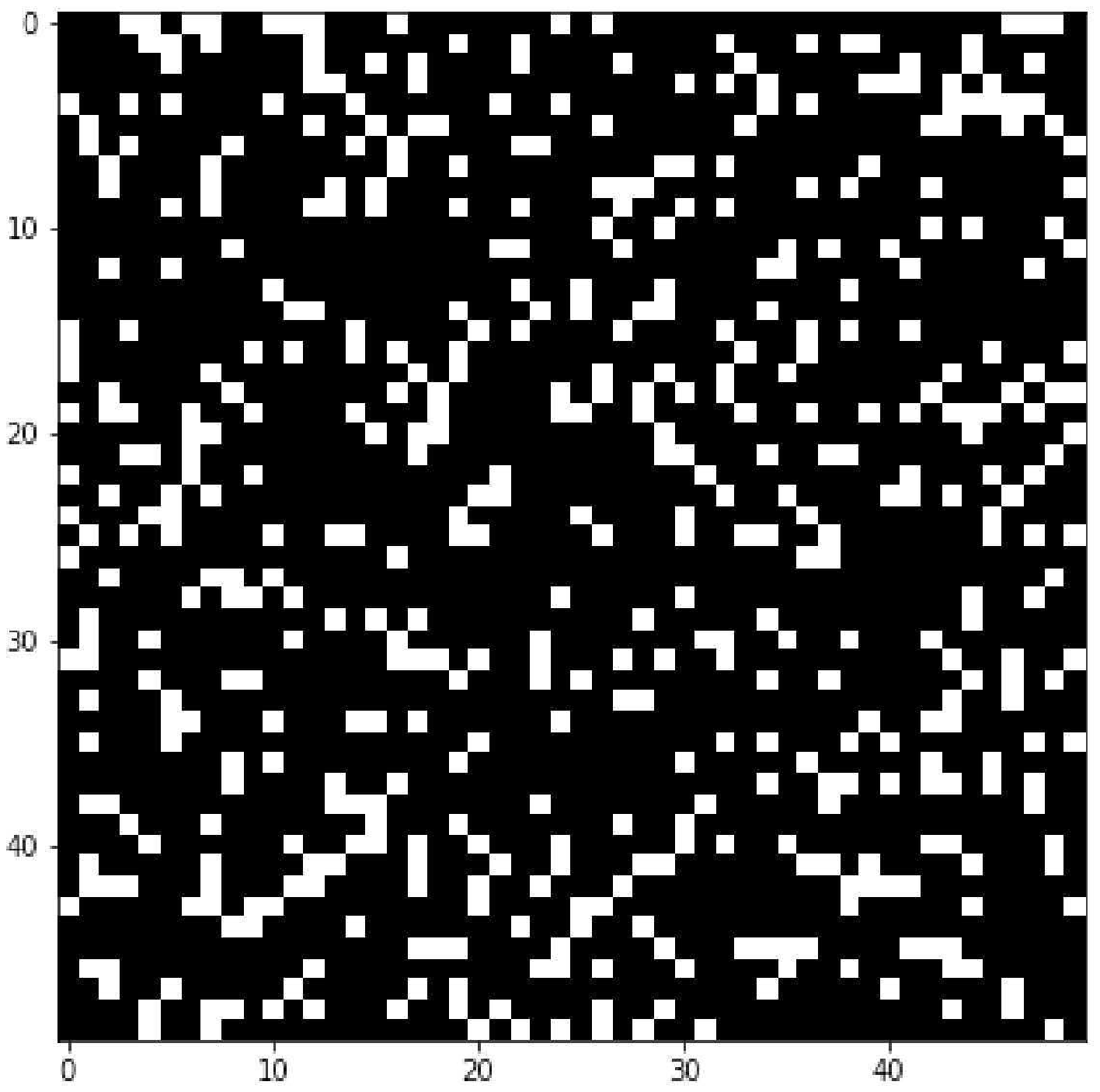}
   \hspace{1.6cm} (c)
     \end{center}
 \end{minipage}\\
 
      \begin{minipage}{0.33\hsize}
  \begin{center}
   \includegraphics[width=42mm]{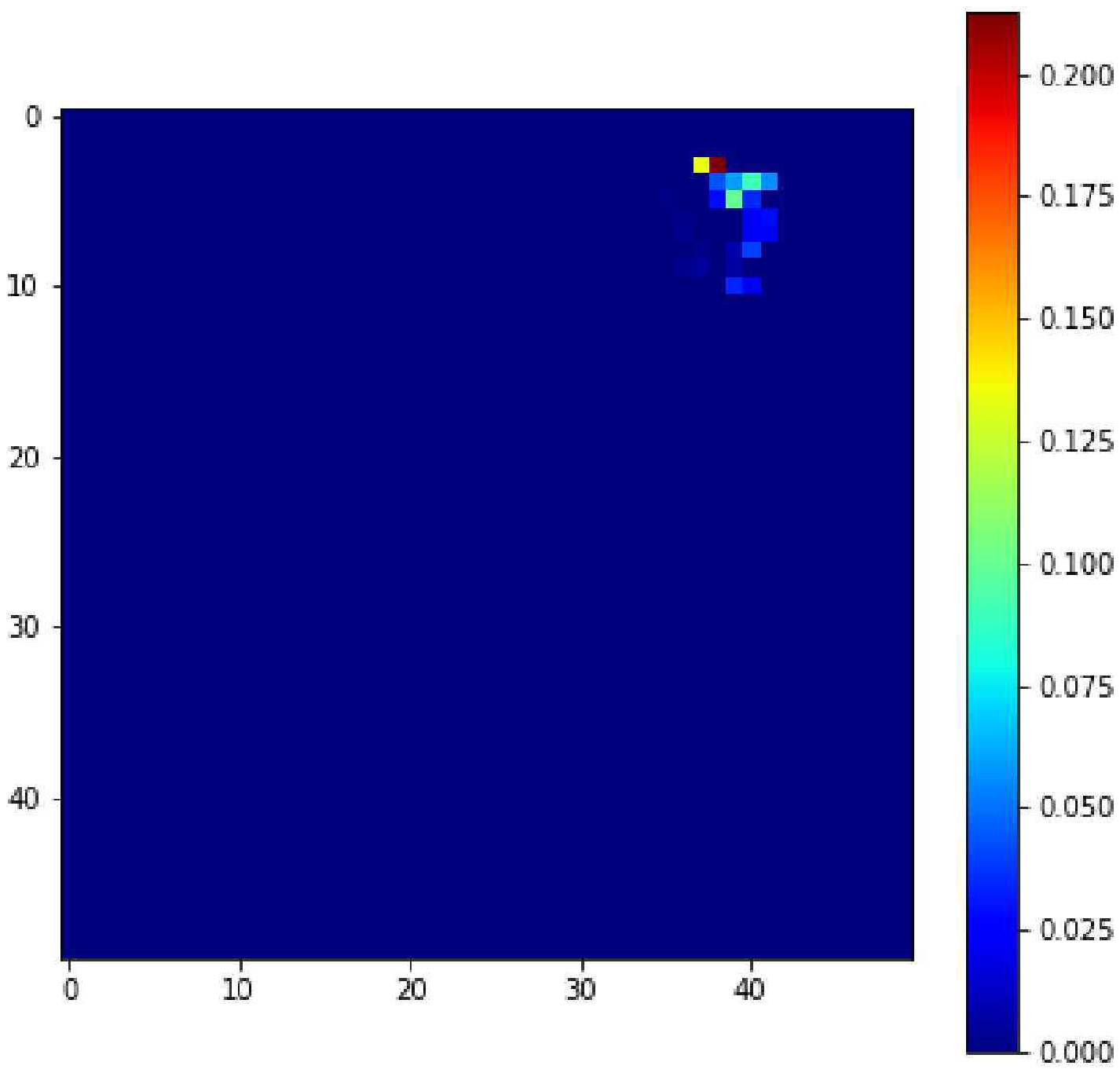}
   \hspace{1.5cm} (d)
     \end{center}
 \end{minipage}

     \begin{minipage}{0.33\hsize}
  \begin{center}
   \includegraphics[width=42mm]{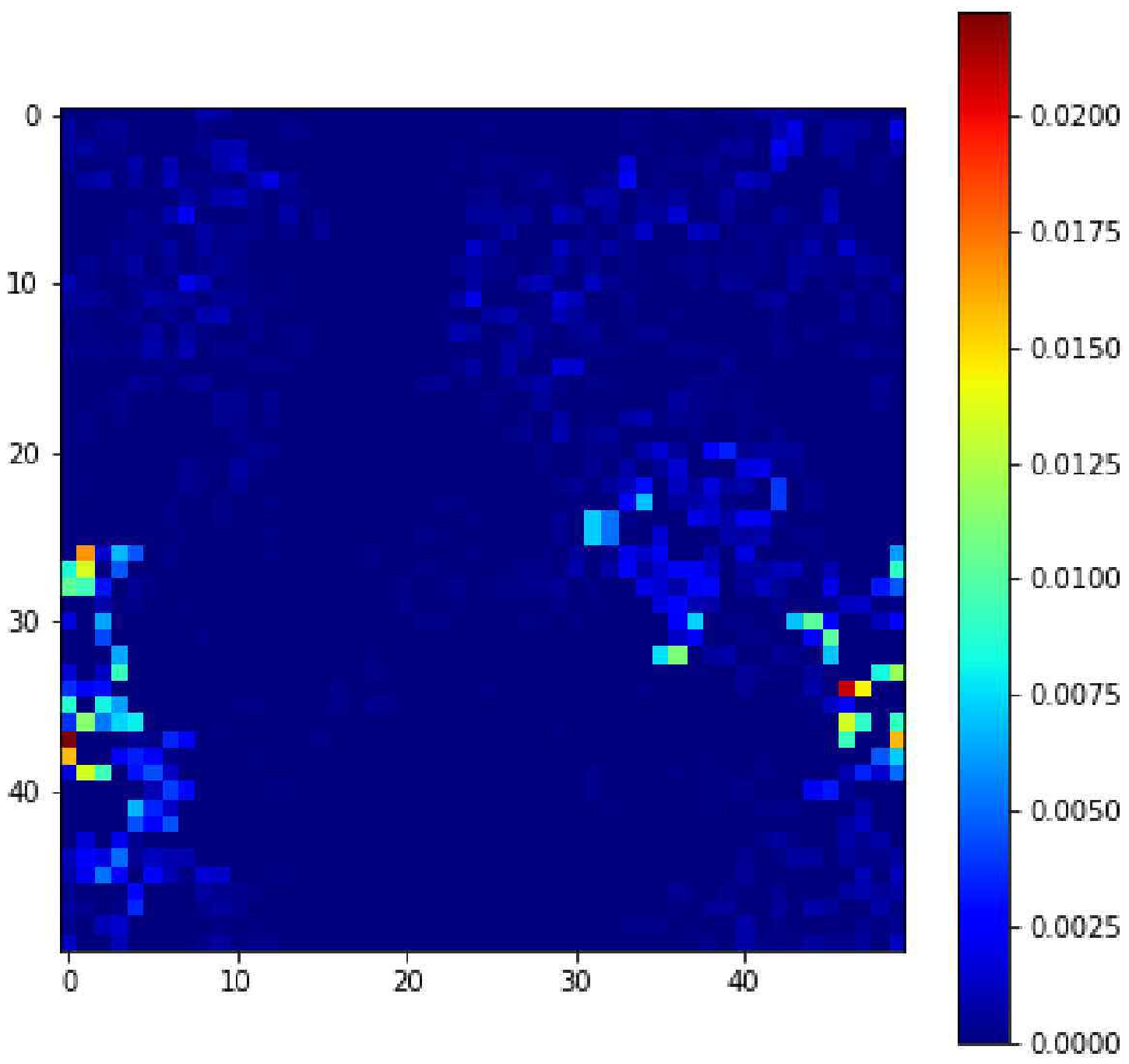}
   \hspace{1.5cm} (e)
     \end{center}
 \end{minipage}

     \begin{minipage}{0.33\hsize}
  \begin{center}
   \includegraphics[width=42mm]{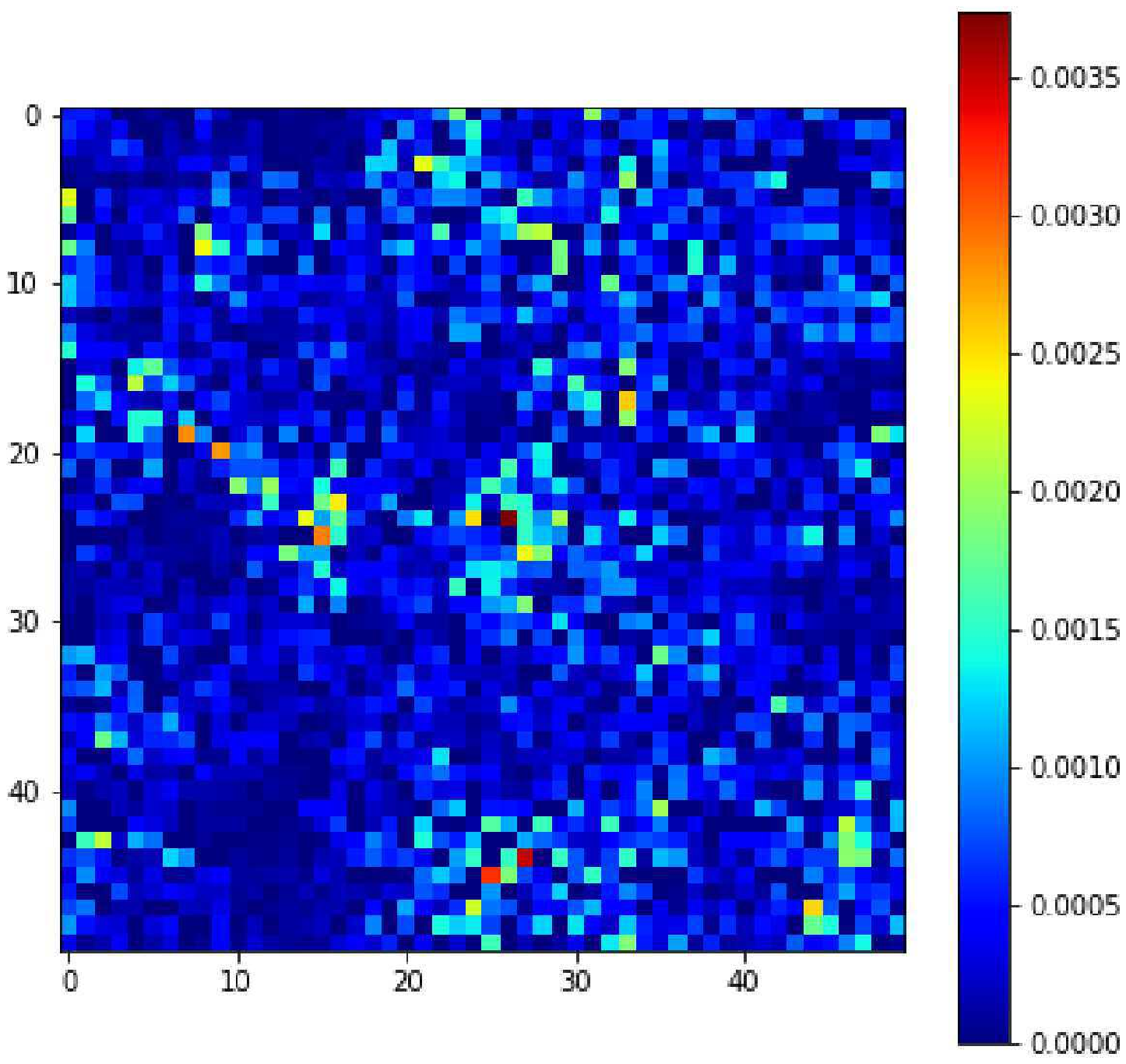}
   \hspace{1.5cm} (f)
     \end{center}
 \end{minipage}

  \end{tabular}
\caption{(Color online) Schematic of 2D site percolation problem, where black cells indicate occupied sites.
 (a) For the site occupation probability $p_\mathrm{s}=0.38<p_{\rm c}=0.5927\cdots$, even the largest cluster does not extend over the system.
(b) Once the probability is above $p_{\rm c}$, for example, $p_\mathrm{s}=0.62>p_{\rm c}$, a cluster connecting one edge
to the opposite edge is formed.
(c) Well above the percolation threshold, $p_\mathrm{s}=0.81>p_{\rm c}$, most of the sites belong to the largest cluster.
(d)--(f) Examples of the wave function formed on the cluster.
Even if the cluster extends all over the system, the wave functions remains localized (e).
The wave functions begin to be extended only when $p_\mathrm{s}$ is well above $p_{\rm c}$ (f). 
}
\label{fig:percolation}
\end{figure}

In the case of quantum percolation, we set the nearest neighbors to be connected randomly,
i.e.,  $V_{\bm{x},\bm{x}'}$ is finite or 0 randomly.
In this paper, we consider the site percolation problem, where
the sites are occupied randomly with probability $p$, and they are connected only when
both of the nearest-neighbor sites are occupied.
The connected sites form clusters, and when $p\ge p_{\rm c}$, a cluster that connects one side of the system to the other appears.
This $p_{\rm c}$ is called the classical percolation threshold.
For $p<p_{\rm c}$, all the clusters are isolated, and the wave functions on them cannot extend over the whole system. 
Thus, the system is always an insulator.  The metal phase, however, does not necessarily appear for $p\ge p_{\rm c}$,
because the wave functions on a cluster may remain localized even if the cluster is extended all over the system.
The condition $p>p_{\rm c}$ is, therefore, a necessary but not sufficient condition. 
 Only when $p\ge p_\mathrm{q}\ge p_{\rm c}$ does
the current flow, where $p_\mathrm{q}$ is the quantum percolation threshold.
See Fig.~\ref{fig:percolation}, where cases of 2D site percolation, the percolation threshold
$p_{\rm c}$ of which is $0.5927\cdots$,
are shown.  Note that all the states are localized in two dimensions\cite{Abrahams79}
except for the symplectic class.  We therefore assumed SU(2) transfer [see Eq.~(\ref{eq:su2})] to observe the localization--delocalization transition
in the case of the quantum percolation problem [Figs.~\ref{fig:percolation}(d)--\ref{fig:percolation}(f)].

In the following subsections, we draw various phase diagrams for many types of disordered system
using the CNN.

\subsection{3D Anderson model and quantum percolation}
We first consider a 3D Anderson model of localization,\cite{Anderson58}
\begin{equation}
\label{eq:Hamiltonian3DAM}
H=\sum_{\bm{x}}
v_{\bm{x}} c_{\bm{x}}^\dagger c_{\bm{x}}-
\sum_{\langle \bm{x},\bm{x}'\rangle} 
c_{\bm{x}}^\dagger c_{{\bm{x}}'}\,,
\end{equation}
where $v_{\bm{x}}$ is randomly and uniformly distributed in the range $[-W/2,W/2 ]$,
with $W$ the disorder strength.
Conventional notations use ``$W$" for the weight
parameters and the strength of disorder.  In this paper, we follow this convention, but to avoid confusion,
the weight parameters are
written as vectors or with indices such as in Eqs.~(\ref{eq:conv}), (\ref{eq:conv2}), and (\ref{eq:finalPhase}), whereas the disorder strength is written as a scalar.

At energy $E=0$, i.e., at the center of the band, the wave functions are delocalized
when $W<W_c$ and the system is a  metal.  For $W>W_c$, the wave functions
are exponentially localized and the system is an Anderson insulator (AI).
Here, the critical disorder $W_c$ is estimated to be $16.54\pm 0.01$ by the finite size
scaling analysis of the Lyapunov exponent calculated by the transfer matrix
method.\cite{Slevin14,Slevin18}

{\it 3D quantum percolation model.}--
We next consider the 3D quantum site percolation model described by the following Hamiltonian\cite{Avishai92,
Berkovits96,Kaneko99,Ujfalusi14},
\begin{equation}
\label{eq:percolationHamiltonian}
H=
-\sum_{\langle \bm{x},\bm{x}'\rangle} 
 c_{\bm{x}}^\dagger \, V_{\bm{x},\bm{x}'} \, c_{{\bm{x}}'}\,,
\end{equation}
where the transfer $V_{\bm{x},\bm{x}'}$ is defined as
\begin{equation}
\label{eq:transfer}
V_{\bm{x},\bm{x}'}=\left\{
\begin{array}{ll}
   1   &  {\rm for\;connected\;bond},\\
   0   &  {\rm for\;disconnected\;bond}.
\end{array}
\right.
\end{equation}
We take the energy unit to be the absolute value of the transfer energy between connected bonds.
In the present case of site percolation, each site is filled with a probability $p_{\rm s}$, and
a bond is connected only when both of the nearest-neighbor sites are filled.
For each realization of site percolations, we identify the maximally connected cluster and analyze
the states on this maximally connected cluster.

According to the study of quantum percolation\cite{Kirkpatrick72,Ujfalusi14}, the strongly localized states, so-called molecular states,
exist at some energies, such as  $E=0, \pm 1, \pm \sqrt{2}, \pm \sqrt{3}$.
These states are peculiar to the quantum percolation model and are degenerate, resulting in the strong peaks in the density of states.
Owing to the degeneracy, any linear combination is possible, which may result in a difficulty of judging the delocalized/localized phases.
We therefore assume a weak site random potential  on the order of $10^{-3}$, namely, we add
$\sum_{\bm{x}}v_{\bm{x}} c_{\bm{x}}^\dagger c_{\bm{x}}$ to the Hamiltonian with $v_{\bm{x}}\in [-10^{-3}/2, 10^{-3}/2]$ and
lift the degeneracy.

In both Anderson and quantum percolation models, 
we consider a $40\times 40\times 40$ simple cubic lattice and impose a periodic boundary condition.
The maximum modulus of the eigenfunction is shifted to the center of the system to improve the accuracy of the machine learning.

For training the neural network, we vary $W$ in the Anderson model in the range $W\in [14,16]$, where $W<W_c= 16.54\pm 0.01$
(metal phase where the wave function is delocalized), and for each $W$, we diagonalize the Hamiltonian via sparse matrix diagonalization
and obtain the eigenfunction closest to the band center, $E=0$.   We choose 4,000 different $W$ to prepare 4,000 eigenfunctions
and label them ``delocalized".  We then change the range of $W$ to $W\in [17,19]$ (insulating phase where the wave function is localized),
prepare 4,000 eigenfunctions, and label them ``localized."
We then set $u_{i}^{(0)}=|\psi(\bm{x}_i)|^2$ and feed these to the 3D CNN and train the neural network so that it recognizes
the localized and delocalized states with high accuracy, typically $>99\%$.

\begin{figure}[htb]
  \begin{center}
     \begin{tabular}{cc}    
 
 \begin{minipage}{0.45\hsize}
  \begin{center}
   \includegraphics[angle=0,width=\textwidth]{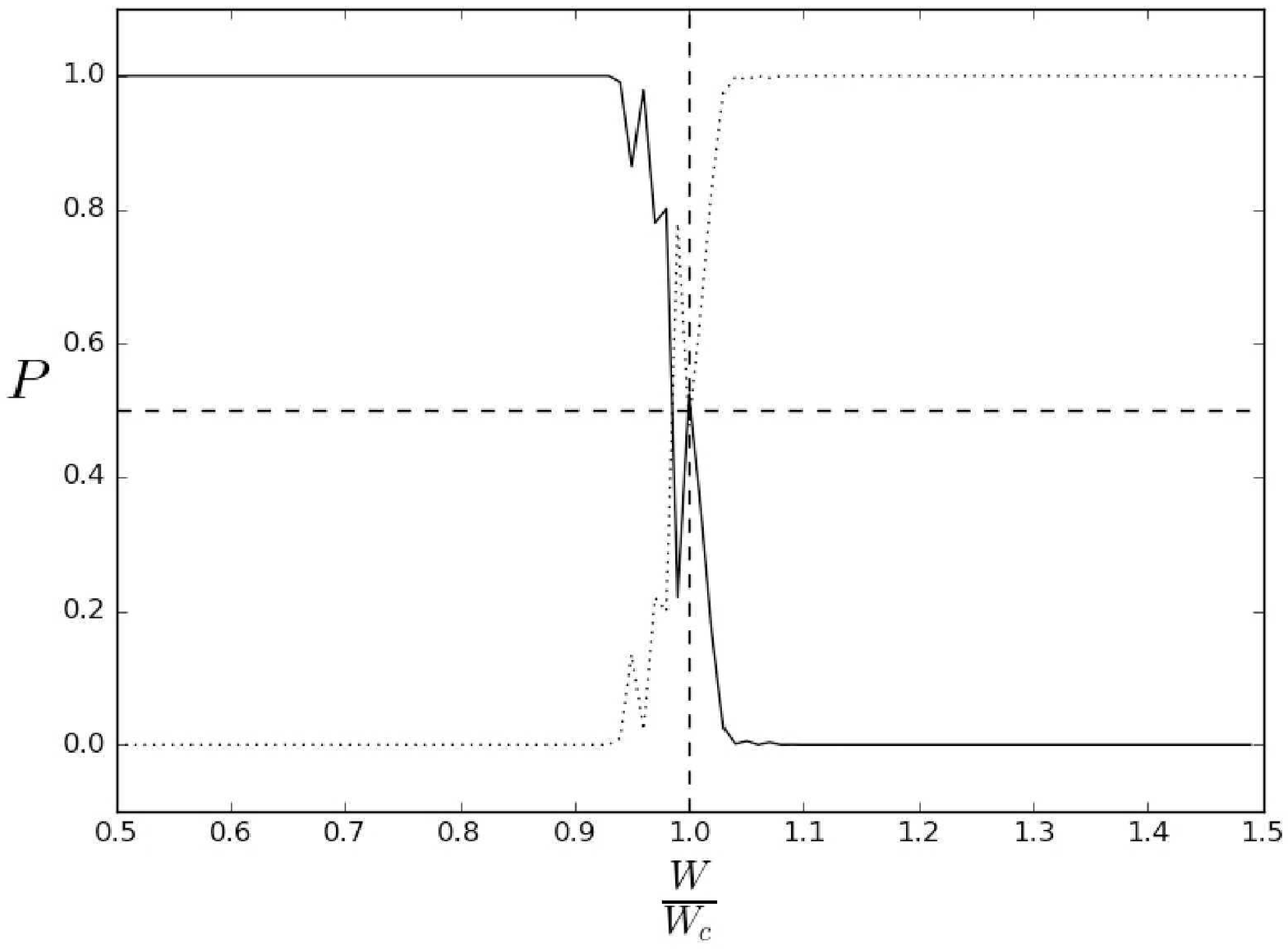}
      \hspace{1.6cm} (a)
     \end{center}
 \end{minipage}
 \begin{minipage}{0.45\hsize}
  \begin{center}
  \includegraphics[angle=0,width=\textwidth]{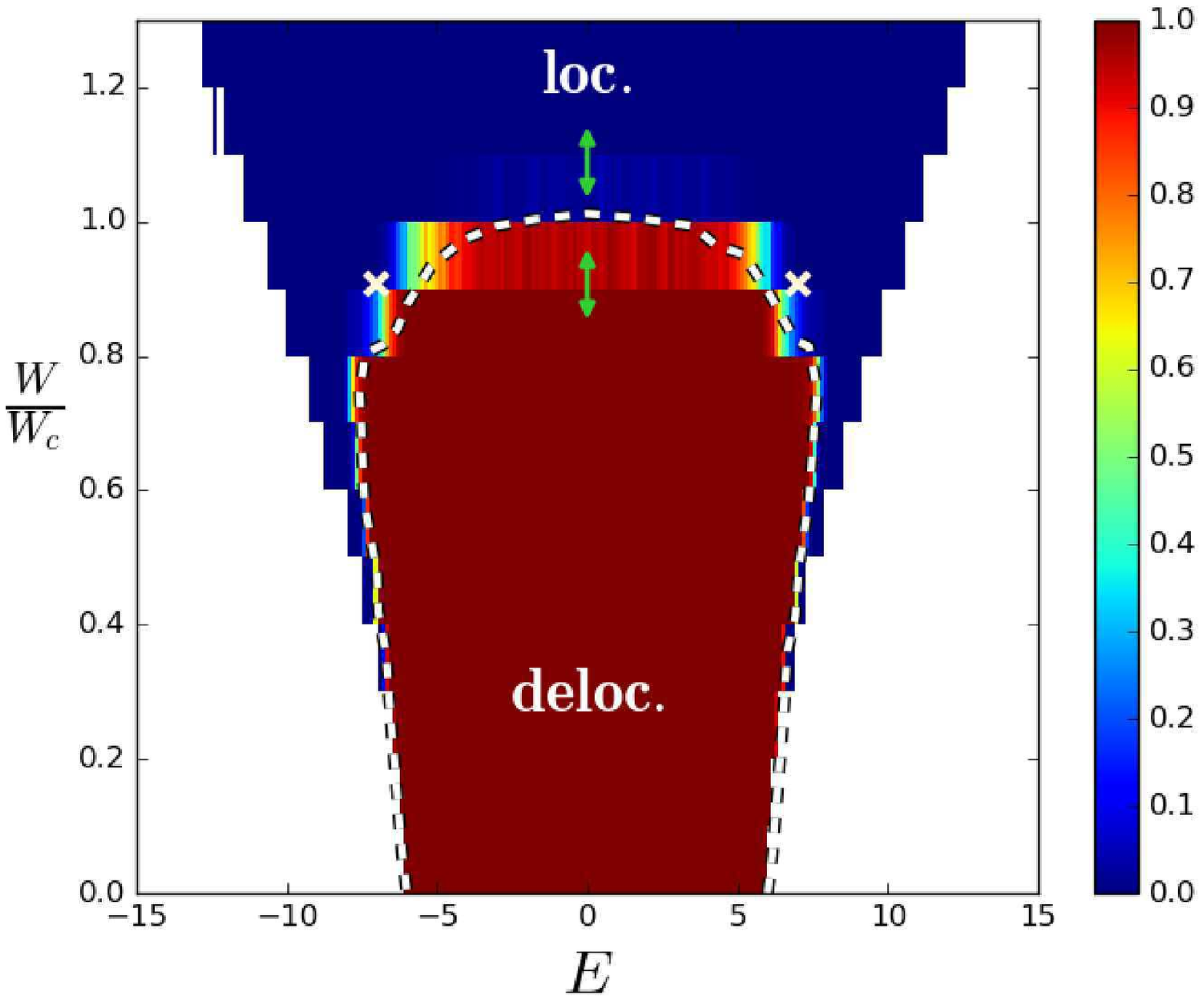}
      \hspace{1.6cm} (b)
     \end{center}
 \end{minipage}
  \end{tabular}

 \caption{(Color online) (a) Probability $P$ that the eigenstates
  are localized/delocalized obtained by the trained CNN
 at $E=0$. (b) Color plot, a type of ``heat map", of $P_\mathrm{deloc}$ in $W$ vs $E$ plane. In (a), the dotted line is $P$ for ``localized" and
 the solid line $P$ for ``delocalized."  The two dashed lines are  guides for
  the eyes and indicate $P=0.5$ and $W=W_c$.
 The eigenstates are prepared independently of the training set.
 The green arrows in (b) indicate the regions where the CNN is trained, whereas the white dashed line and crosses indicate the phase boundary
 estimated by other methods.\cite{Queiroz01,Schubert05}
 Note the appearance of the localized phase near the band edges for small $W$, which is correctly captured
 by the CNN.  Taken from Ref.~\cite{Mano17}.}
\label{fig:AT}
\end{center}
\end{figure}

{\it Result for Anderson model.--}
In Fig.~\ref{fig:AT}(a), we show the probability that the states are localized at $E=0$,
$E$ being the Fermi energy.
The test data is 100 eigenfunctions with various $W$, and the average over five samples is taken as in Fig.~1 of Ref. \cite{Tomoki17}.
From this figure, the phase transition from a delocalized to a localized phase has been confirmed around $W_c$,
from which we confirm that the CNN correctly detected the Anderson transition. 

We then prepare eigenfunctions throughout the energy spectrum with varying $W$,
and let the machine (CNN) determine the phase.
In Fig.~\ref{fig:AT}(b), we plot $0\times P_\mathrm{loc}+1\times P_\mathrm{deloc}=P_\mathrm{deloc}$  as a heat map. 
The sharp change in color from red to blue indicates that the CNN correctly identified 
the metal--insulator transition.
This rapid change in color indicates the phase boundary, which is in good agreement with the previous results.\cite{Bulka87,Queiroz01,Schubert05}

Near the band edges, even for small $W\approx 0.5 W_c$, the machine judged that the eigenstates are localized. These states near the band edges are localized because of potential localization with little quantum interference. 
We note that the CNN has been trained only with the eigenfunctions around $E=0$,
where the localization is caused by quantum interference due to multiple scatterings,
not by potential localization. 

{\it Application to quantum percolation.--}
Now, we apply the CNN trained for the Anderson model at the band center [green arrows in Fig.~\ref{fig:AT}(b)]
to obtain the phase diagrams of the 3D quantum percolation.
Owing to the random connection between the sites, the transfer matrix method is not applicable  [see Eq.~(\ref{eq:tmm})].  In addition, the density of states is spiky for quantum percolation.
Drawing the phase diagram, therefore, is more difficult than in the case of the Anderson model.

\begin{figure}[htb]
  \begin{center}
 \includegraphics[angle=0,width=0.5\textwidth]{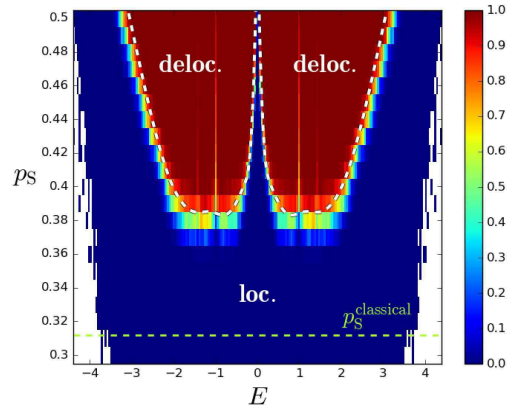}
 \caption{(Color online) Phase diagrams for  site-type quantum percolation
drawn by CNN trained using the Anderson model with box-type site random potential.  The white dashed line
 is from the estimates by Ref.\cite{Ujfalusi14},
 whereas the green horizontal dashed line indicates the classical percolation
 threshold.  Taken from Ref.\cite{Mano17}.}
\label{fig:QP}
\end{center}
\end{figure}

Figure~\ref{fig:QP}  shows the phase diagrams of the 3D site-type quantum percolation model.
The horizontal axis is the eigenenergy, and the vertical axis is the
probability of site occupation, $p_\mathrm{s}$.  The green dashed line indicates the
classical percolation threshold, $p_\mathrm{s}^\mathrm{classical}\approx 0.312$, above which the sites
percolate.\cite{Sur76,Wang13}
We see that the quantum percolation transition occurs well above $0.312$,
which means that even if the sites percolate, the wave function on the sites remains
 localized.  The quantum percolation threshold $p_\mathrm{q}$ depends on the
energy nonmonotonically.
We emphasize that the CNN used to draw this phase diagram is trained in a
small region of the phase diagram, indicated by the green arrows in Fig.~\ref{fig:AT}(b), with no additional training
for quantum percolation.

\subsection{Generalization capability}
As we have seen above, once the CNN is trained in small regions of the phase diagram, 
it can determine the phase outside the training region [Fig.~\ref{fig:AT}(b)]
as well as the phase of a different model such as the quantum percolation model (Fig.~\ref{fig:QP}).
Thus, we have demonstrated the generalization capability of the CNN.

\begin{figure}[htb]
  \begin{center}
     \begin{tabular}{cc}    
 
 \begin{minipage}{0.45\hsize}
  \begin{center}
   \includegraphics[angle=0,width=\textwidth]{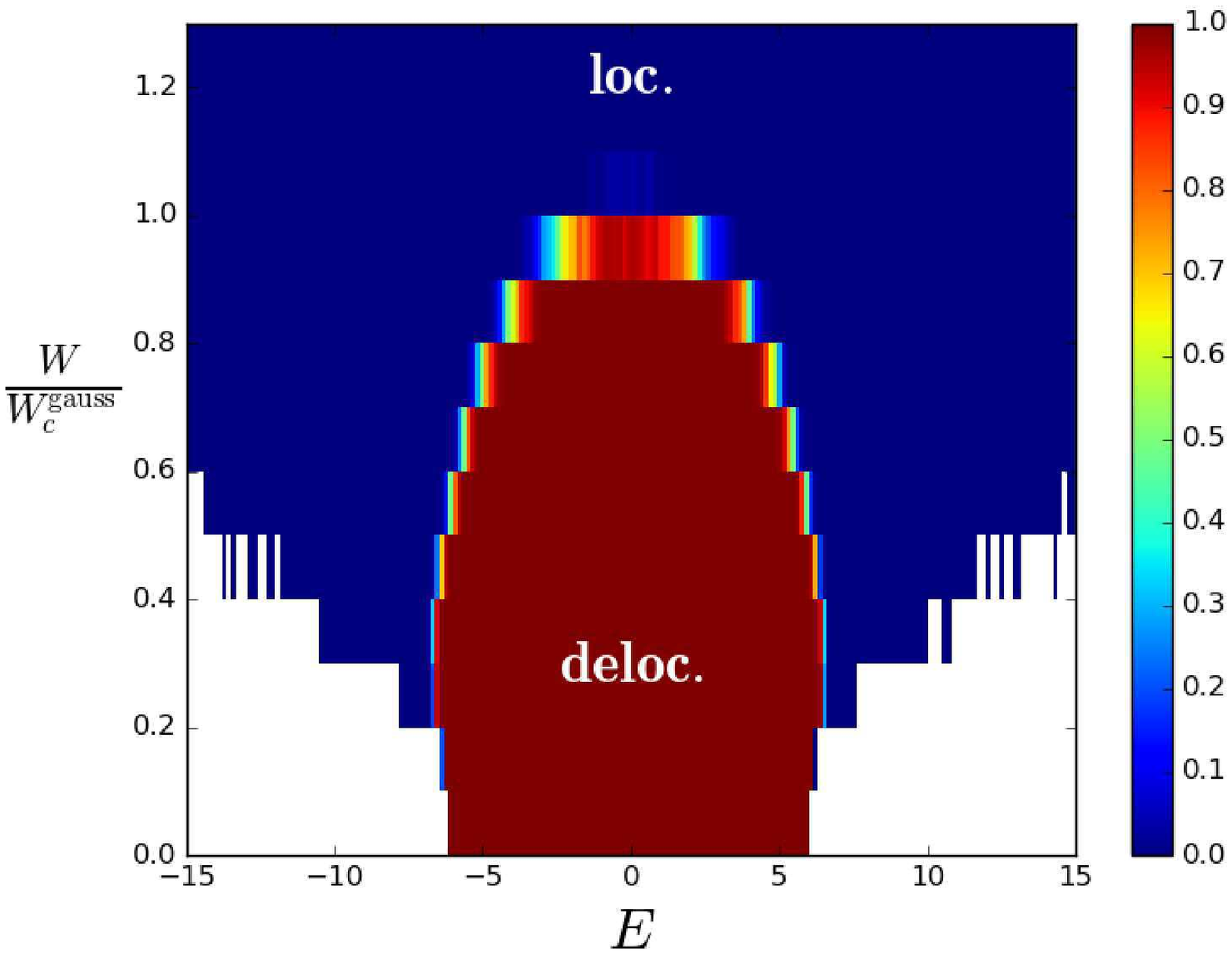}
      \hspace{1.6cm} (a)
     \end{center}
 \end{minipage}
 \begin{minipage}{0.45\hsize}
  \begin{center}
  \includegraphics[angle=0,width=\textwidth]{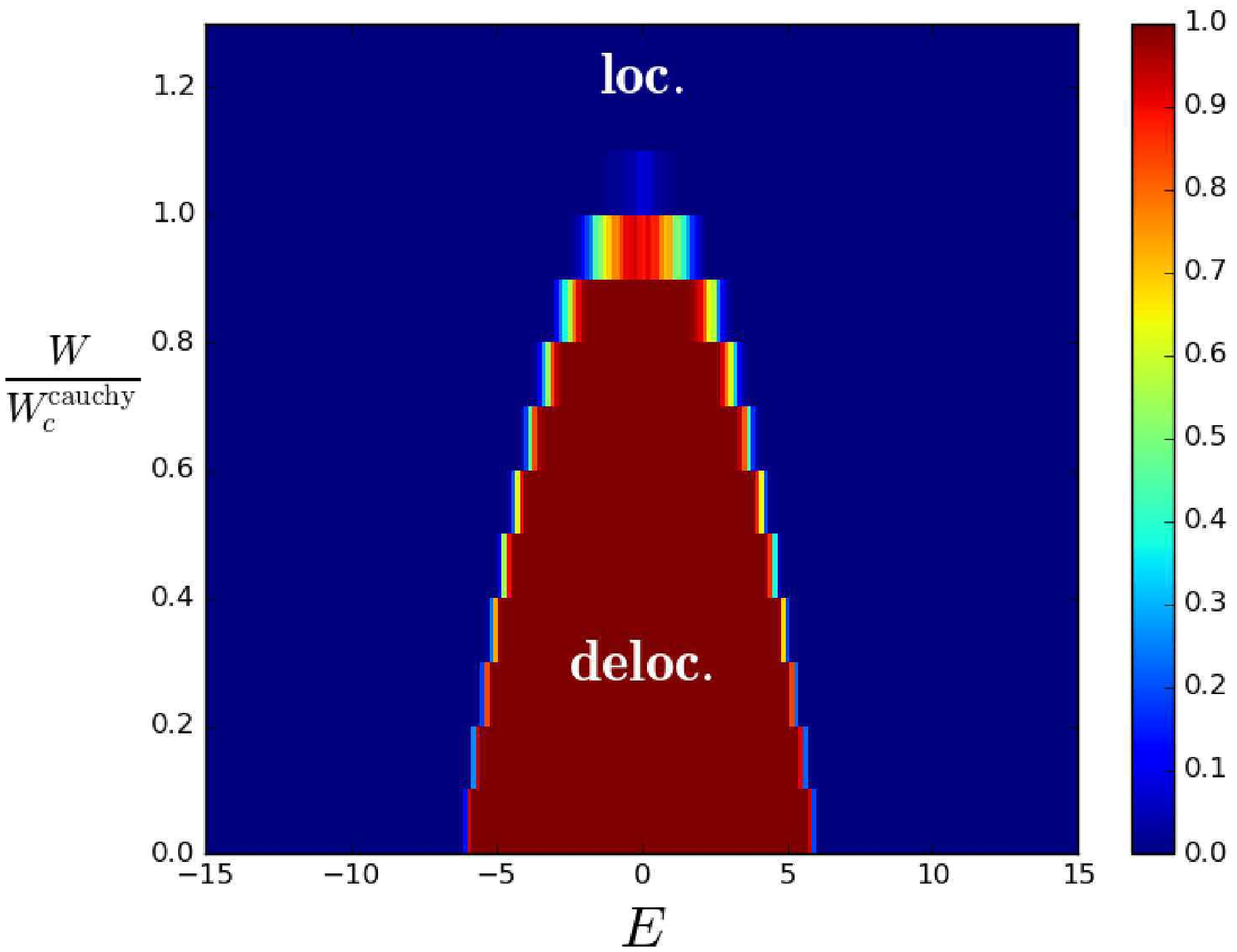}
      \hspace{1.6cm} (b)
     \end{center}
 \end{minipage}
  \end{tabular}

 \caption{(Color online) Same as Fig.~\ref{fig:AT}(b), but with site potential distributions
 Gaussian (a) and Cauchy (b). 
 These phase diagrams are drawn using the CNN trained with the Anderson model with
 a box-type site potential distribution.
 The vertical axes are scaled by
(a) $W_c^\mathrm{gauss}\approx 6.147$ and  (b) $W_c^\mathrm{cauchy}\approx 4.2707$.\cite{Slevin14}
}
\label{fig:ATGaussCauchy}
\end{center}
\end{figure}

We can further test the generalization capability by changing the site potential
of the Anderson model from the random box distribution, whose
probability distribution is $P(v_{\bm{x}})=\frac{1}{W}\Theta\left(\frac{W}{2}-|v_{\bm{x}}|\right)$,
 to Gaussian and Cauchy distributions,
$P(v_{\bm{x}})=\frac{1}{\sqrt{2\pi W^2}} \exp\left(-\frac{v_{\bm{x}}^2}{2W^2}\right)$ and 
$P(v_{\bm{x}})=\frac{W}{\pi(v_{\bm{x}}^2+W^2)}$, respectively.  See Fig.~\ref{fig:ATGaussCauchy}.
From Figs.~\ref{fig:AT}(b) and \ref{fig:ATGaussCauchy}(a),
we note that near the band edges, if we fix $E$ as, say, 6.5, and increase $W$,
the insulator--metal--insulator transition occurs for the cases of box and Gaussian distributions,
which is known as reentrant behavior of the Anderson transition.\cite{Bulka87}
Such reentrant behavior does not show up in the case of the
 Cauchy distribution [Fig.~\ref{fig:ATGaussCauchy}(b)].

\begin{figure}[htb]
  \begin{center}
 \includegraphics[angle=0,width=0.5\textwidth]{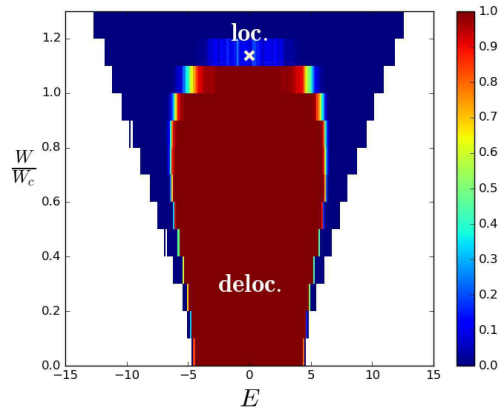}
 \caption{(Color online) Phase diagram of the Anderson transition with broken TRS.
 The cross $\times$ indicates
the critical point $W\approx 18.8$ estimated by the transfer matrix method.\cite{Kawarabayashi98,Slevin16}
 The disorder strength, $W$, is scaled by the critical disorder of the orthogonal
class, $W_c=16.54$.  The phase at $(E, W)=(0,1.05 W_c)$ is an insulator in Fig.~\ref{fig:AT}(b),
whereas it is a metal in the present case, indicating magnetic-field-induced delocalization.}
\label{fig:ATUnitary}
\end{center}
\end{figure}

We can also break the TRS by adding random phases
to the transfer, $V_{\bm{x},\bm{x}'}=\exp(i\theta_{\bm{x},\bm{x}'})$,
with $\theta_{\bm{x},\bm{x}'}$ uniformly distributed in $[0,2\pi)$.\cite{Kawarabayashi98}
Figure~\ref{fig:ATUnitary} shows the phase diagram.  The cross $\times$ indicates
the estimate by the transfer matrix method \cite{Slevin16}, which
is consistent with the present results.
To compare the phase diagram of the orthogonal class, Fig.~\ref{fig:AT}(b),
we scale the disorder strength, $W$, by the critical disorder of the orthogonal
class, $W_c=16.54$.
The states at $E=0, W/W_c=1.05$ are localized in the orthogonal class [Fig.~\ref{fig:AT}(b)],
but are delocalized in Fig.~\ref{fig:ATUnitary}.
This is contrary to the naive expectation that the addition of the random phases results in stronger disorder;
hence, the localization is enhanced, not suppressed.
In fact, the effect of breaking TRS, which causes delocalization,
overcomes the effect of the addition of randomness, leading to
random-magnetic-field-induced delocalization.
This nontrivial feature of the phase diagram is correctly captured by the CNN.

\subsection{2D SU(2) model and quantum percolation}
\label{sec:2dsu2}
In random non-interacting electron systems,
all the states are localized in two dimensions and there are no metal phases.\cite{Abrahams79}
In the presence of spin--orbit scattering, however, electrons can be extended,\cite{Hikami80,Efetov80, Hikami81}
and the system undergoes a metal--insulator transition with change in the disorder
strength and Fermi energy.

To incorporate the spin--orbit interaction in the tight-binding model, Eq. (\ref{eq:Hamiltonian}),
we choose the transfer $V_{\bm{x},\bm{x}'}$ to be SU(2) matrices.\cite{Evangelou87,Ando89,Asada02}
To analyze the localization--delocalization transition, which is characterized by the divergence of localization/correlation lengths $\xi$,
other length scales such as the spin--precession length should be much shorter than $\xi$.
We therefore take $V_{\bm{x},\bm{x}'}$ to be random.\cite{Evangelou87,Asada02}
Of all the choices of the probability distribution of $V_{\bm{x},\bm{x}'}$,
we take the invariant Haar measure,
\begin{equation}
V_{\bm{x},\bm{x}'}=
\begin{pmatrix}
e^{i\alpha_{\bm{x},\bm{x}'}}\cos\beta_{\bm{x},\bm{x}'} & e^{i\gamma_{\bm{x},\bm{x}'}}\sin\beta_{\bm{x},\bm{x}'} \\
-e^{-i\gamma_{\bm{x},\bm{x}'}}\sin\beta_{\bm{x},\bm{x}'} & e^{-i\alpha_{\bm{x},\bm{x}'}}\cos\beta_{\bm{x},\bm{x}'}
\end{pmatrix}\,,
\label{eq:su2}
\end{equation}
with $\alpha$ and $\gamma$  uniformly distributed in the range $[0,2\pi)$.
The probability density $P(\beta)$ is
\begin{equation}
P(\beta)=\left\{
\begin{array}{ll}
 \sin(2\beta)     &  0\le \beta\le \pi/2  \,,\\
   0   &   {\rm otherwise}.
\end{array}\right.
\label{eq:su2_2}
\end{equation}

\begin{figure}[htb]
  \begin{center}
 \includegraphics[angle=0,width=0.5\textwidth]{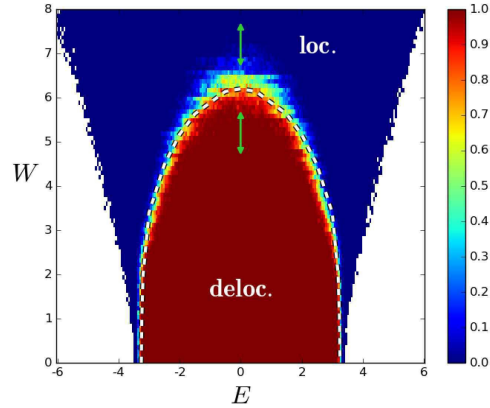}
 \caption{(Color online) Heat map plot of $P_\mathrm{deloc}$ in $W$ vs $E$ plane
 for the 2D Anderson transition with TRS but with
 broken  SRS (symplectic universality class).
The system size is $40\times 40$, where we prepared 4,000 2D wave functions in the metal phase
 and 4,000 in the insulating phase.
 The arrows are the regions used for supervised training.
The dashed line is the estimate of the transfer matrix method.\cite{Asada04}}
\label{fig:2d_su2}
\end{center}
\end{figure}

The Anderson transition of this 2D system is well detected by the CNN. 
In Fig.~\ref{fig:2d_su2}, we plot the probability $P_\mathrm{deloc}$.
The sharp change in the color at the metal-insulator phase boundary (dashed line\cite{Asada04})
indicates that the CNN has correctly detected the Anderson transition.

\begin{figure}[htb]
  \begin{center}
 \includegraphics[angle=0,width=0.5\textwidth]{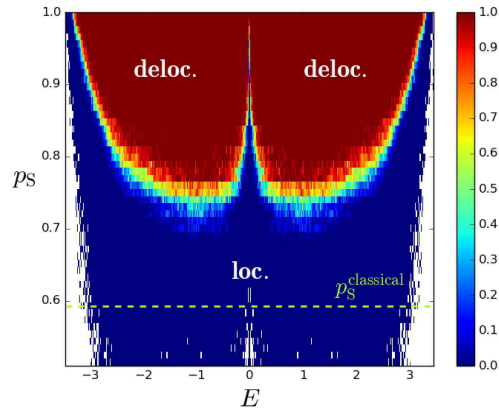}
 \caption{(Color online) Phase diagrams for  site-type 2D quantum percolation with random SU(2) transfer
drawn using the CNN trained with the 2D Anderson model with box-type site random potential.
 The dashed lines indicate the classical percolation threshold.}
\label{fig:2d_su2_percolation}
\end{center}
\end{figure}

As in the case of the 3D Anderson transition, once the CNN is trained for a regular square lattice,
we can apply this CNN to the quantum percolation model where the lattice is random.
Figure~\ref{fig:2d_su2_percolation} shows the phase diagram of the 2D quantum percolation for the symplectic class.
As in the 3D quantum percolation, the 2D quantum percolation threshold is
significantly higher than the classical percolation threshold (dashed line) of $p_\mathrm{s}^\mathrm{classical}\approx 0.5927\cdots$.\cite{StaufferBook}

Note that there is no Anderson transition for the orthogonal class.  In the case of the unitary class,
the quantum Hall transition\cite{Klitzing80,Huckestein95,Chalker88,Kramer05} takes place in high magnetic fields or in a Chern insulator.
The supervised training approach is also valid for this quantum Hall transition,\cite{Tomoki16,Zhang17a}
where the critical exponent is extracted.\cite{Li17}

\subsection{Anderson transition and quantum percolation in higher dimensions}
It is instructive to discuss the Anderson transition and quantum percolation
in higher dimensions.  The critical exponent $\nu$ is known to be $1/2$ in the
limit of infinite dimensions,\cite{Efetov90,Mirlin94,Garcia-Garcia08}
and the Borel-Pad\'e approximation\cite{Hikami92}
successfully interpolates the exponents in low to high dimensions.\cite{Ueoka14,Ueoka17}
The Anderson transition in four dimensions can be studied, for example, using the quantum kicked rotor
with amplitude modulation realized in atomic matter waves.\cite{Chabe08}

\begin{figure}[htb]
  \begin{center}
     \begin{tabular}{cc}    
 
 \begin{minipage}{0.45\hsize}
  \begin{center}
   \includegraphics[angle=0,width=\textwidth]{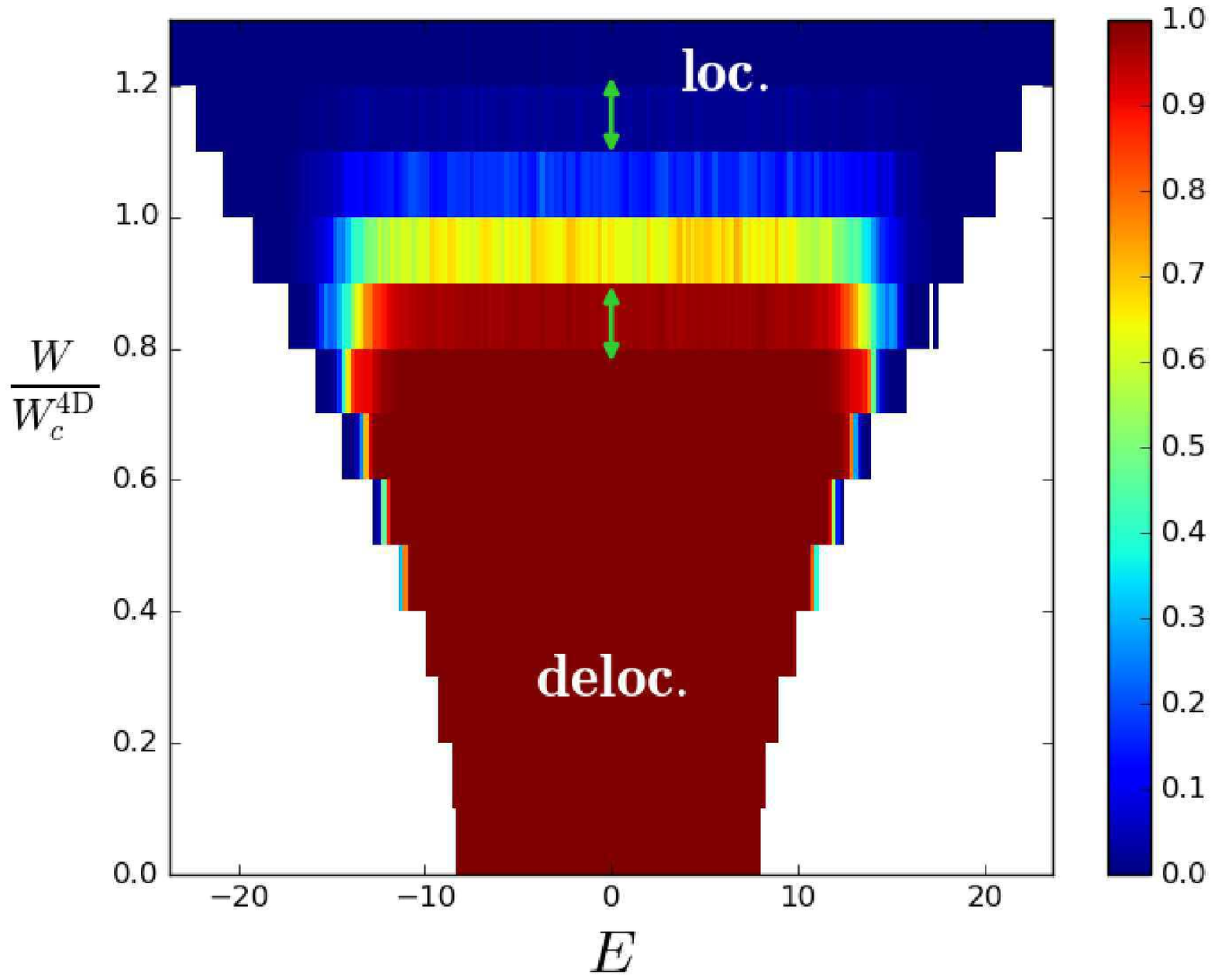}
      \hspace{1.6cm} (a)
     \end{center}
 \end{minipage}
 \begin{minipage}{0.45\hsize}
  \begin{center}
  \includegraphics[angle=0,width=\textwidth]{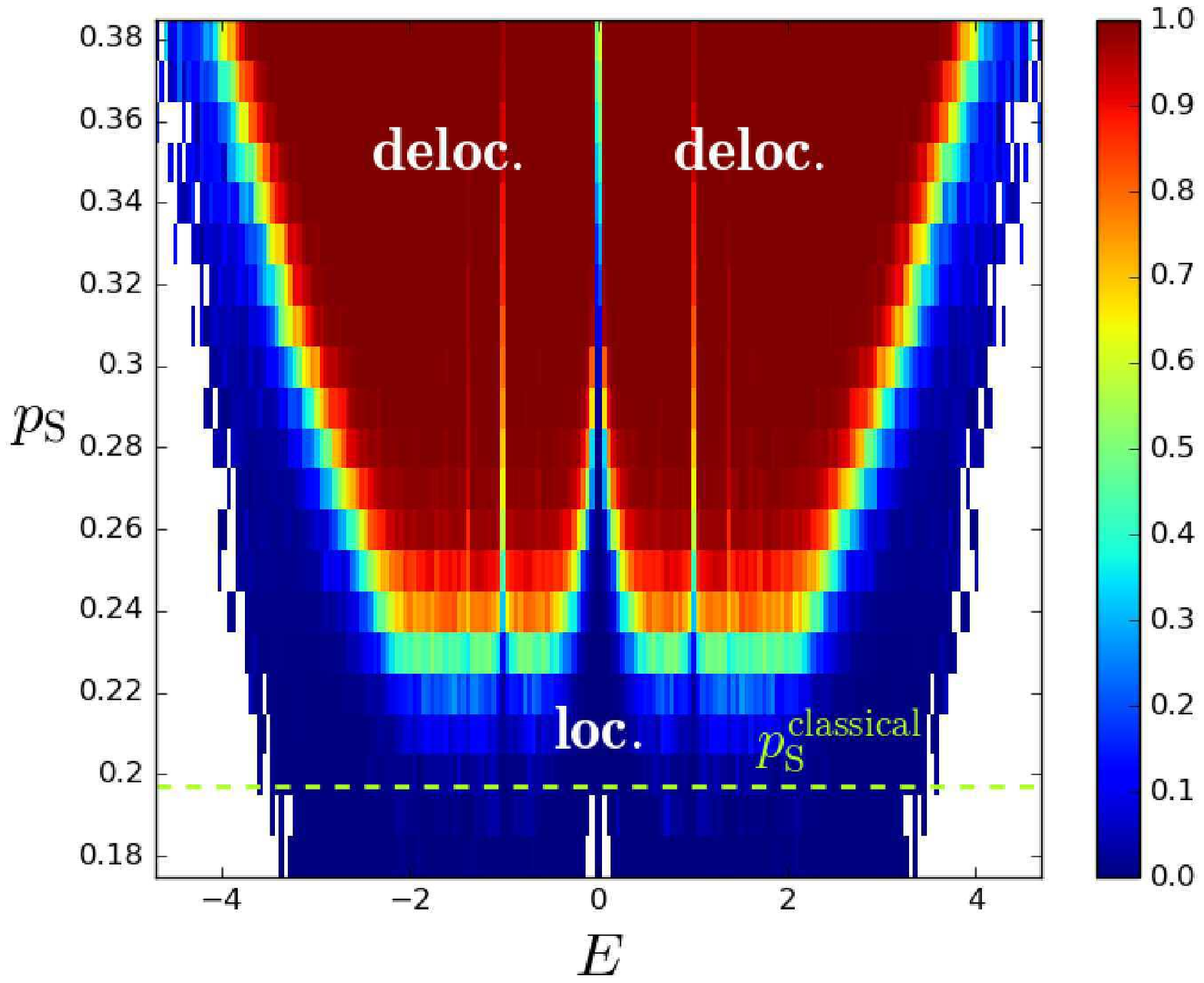}
      \hspace{1.6cm} (b)
     \end{center}
 \end{minipage}
  \end{tabular}

 \caption{(Color online) Phase diagrams for the 4D Anderson transition (a) and 4D quantum percolation (b).
 The green arrows indicate the training region.  In (a), the disorder strength $W$ is scaled by
 $W_c^\mathrm{4D}\approx 34.62$\cite{Ueoka14}.
 In the quantum percolation model, again the
 quantum percolation threshold is well above the classical percolation threshold (dashed line).}
\label{fig:ATQP4d}
\end{center}
\end{figure}

For human beings, the wave functions in 4D space are difficult to imagine and analyze,
since our eyes and brains are already trained to observe 2D and 3D images.
For a machine, it does not matter whether the images are 3D or 4D.
As in the case of the 3D Anderson transition and quantum percolation, we prepare 2,000 4D wave functions in the metal phase ($27<W<31<W_c^\mathrm{4D}\approx 34.62$)
and 2,000 in the insulating phase ($W_c^\mathrm{4D}<38<W<42$)
by diagonalizing the 4D Anderson models [cf. Eq. (\ref{eq:Hamiltonian})] of size
$16\times 16\times 16\times 16$.  Once the CNN is trained, we feed the wave functions of unknown
phases and let the CNN determine the phase.  The results are shown in Fig.~\ref{fig:ATQP4d}(a).
The CNN trained for the 4D Anderson model is then used to draw the phase diagram of
4D quantum site percolation, Fig.~\ref{fig:ATQP4d}(b).
Again, the quantum percolation threshold, i.e., the localization--delocalization phase boundary,
is well above the classical percolation threshold $p_\mathrm{s}^\mathrm{classical}\approx 0.197$.\cite{Mertens18}

\subsection{3D topological matter}
Some of the band insulators are now recognized as topological insulators,\cite{Hasan10,Moore10,Qi11,Ando13,Vergniory19}
where
the bulk wave functions have nontrivial topology.  As a consequence,
the interface between the bulk (nontrivial) and vacuum (trivial) shows edge/surface states.
Another interesting topological material is the 3D Weyl semimetal,\cite{Murakami07,Wan11} where
the hybridization of surface states and bulk Weyl nodes appears.  See Figs. \ref{fig:wfATReal} and
\ref{fig:wfATFourier} in Appendix.
Here, we use the CNN to detect these novel surface states of 3D topological insulators and Weyl
semimetals.
One of the advantages of detecting the surface states is that we can detect the topological
phase even in the presence of randomness, which breaks translational invariance.

\subsubsection{3D topological insulators}
We first consider the topological insulators using the Wilson--Dirac-type tight-binding Hamiltonian,\cite{Liu:3DTI,RyuNomura:3DTI}
   \begin{align} \label{eqn:H}
      H = & \sum_{\bm{x}} \sum_{\mu=x,y,z} \left[\frac{{\rm i}t}{2} c^{\dag}_{{\bm{x}}+{\bf e}_\mu} \alpha_{\mu} c_{\bm{x}}
                                         -\frac{m_{2,\mu}}{2}  c^{\dag}_{{\bm{x}}+{\bf e}_\mu} \, \beta c_{\bm{x}} + \rm{H.c.}\right]  \nonumber \\
            & + (m_0+\sum_{\mu=x,y,z} m_{2,\mu})\sum_{\bm{x}} c^{\dag}_{\bm{x}} \, \beta c_{\bm{x}}
            + \sum_{\bm{x}} v_{\bm{x}} c^{\dag}_{\bm{x}} 1_{4} c_{\bm{x}},
   \end{align}
where $c^{\dag}_{\bm{x}}$ ($c_{\bm{x}}$) is a four-component creation (annihilation) operator on a simple cubic lattice
 at site $\bm{x}$,
and ${\bf e}_{\mu}$ is a unit vector in the $\mu$-direction. $\alpha_{\mu}$ and $\beta$ are gamma matrices defined by
   \begin{align} \label{eqn:gammaMat}
      \alpha_{\mu} =\tau_x\otimes\sigma_\mu= \begin{pmatrix}
                         0     & \sigma_\mu \\
                      \sigma_\mu &    0
                   \end{pmatrix}, \ 
      \beta =\tau_z\otimes 1_2= \begin{pmatrix}
                      1_{2} & 0 \\
                      0 & -1_{2}
                   \end{pmatrix}, 
   \end{align}
where $\sigma_{\mu}$ and $\tau_\mu$ are Pauli matrices that act on the spin and orbital degrees of freedom, respectively.
$m_0$ is the mass parameter, and $m_{2,\mu}$ and $t$ are transfer energies.
In the absence of randomness, the energy band reads
\begin{equation}
\label{eq:tiEnergyband}
E(\bm{k})=\pm \sqrt{m(\bm{k})^2+t^2(\sin^2 k_x+\sin^2 k_y+\sin^2 k_z)}\,,\, m(\bm{k})=m_0+\sum_{\mu=x,y,z}m_{2,\mu}(1-\cos k_\mu)\,.
\end{equation}

The random potential $v_{\bm{x}}$ is uniformly and independently distributed between $[-W/2,W/2]$.
For simplicity, we set $m_{2,x} =m_{2,y}=1$ as the energy unit.
 This Hamiltonian belongs to the symplectic class for $W>0$.
We impose fixed boundary conditions in a direction (in this case, $x$-direction),
so the surface states, if any, appear on surfaces normal to the $x$-direction.
Strong topological insulators (STIs) show 2D Dirac cones on all their surfaces,
one Dirac cone for each surface,
whereas weak topological insulators (WTIs) show even numbers of 2D Dirac cones for each surface.
The WTI is further characterized by weak index $(i,j,k)$, which means that no
surface states appear normal to the $(i,j,k)$-direction.

Systems of size $24\times 24\times 24$ are diagonalized numerically,
and the state whose eigenenergy is closest to the band center $E=0$ is taken.
In the following, we set $t=2$ and $m_{2,z}=0.5$.  In this case, the ordinary insulator (OI)
phase appears in $m_0>0$,
the STI phase in $0>m_0>-1$,
the WTI phase with weak index (001)  in $-1>m_0>-2$,
and  the WTI phase with weak index (111)  in $-2>m_0>-3$.
\cite{Imura12,Kobayashi15}

The eigenfunctions for the state $|\nu\rangle$ have four components due to spin and orbital degrees
of freedom, and are denoted as
$\psi_\nu(x,y,z,i)\,,(i=1,2,3,4)$.
We define a 3D image by $|\psi_\nu(x,y,z)|^2 = \sum_i^4 |\psi_\nu(x,y,z,i)|^2$.
In Ref.\cite{Tomoki17}, the 3D wave function is mapped to a 2D image by
integrating $|\psi(x,y,z)|^2$ over one direction, for example, the $z$-direction,
\begin{equation}
\label{eq:2dTo3d}
F(x,y)=\int \mathrm{d}z\, |\psi(x,y,z)|^2\,,
\end{equation}
and the surface states that extend parallel to the $z$-direction become edge states in the 2D image.
This method, however, has difficulty in distinguishing STI from WTI(111).
In this paper, we use 3D image recognition to distinguish these different topological phases.

\begin{figure}[htb]
  \begin{center}
     \begin{tabular}{cc}     
 \begin{minipage}{0.48\hsize}
  \begin{center}
   \includegraphics[angle=0,width=0.9\textwidth]{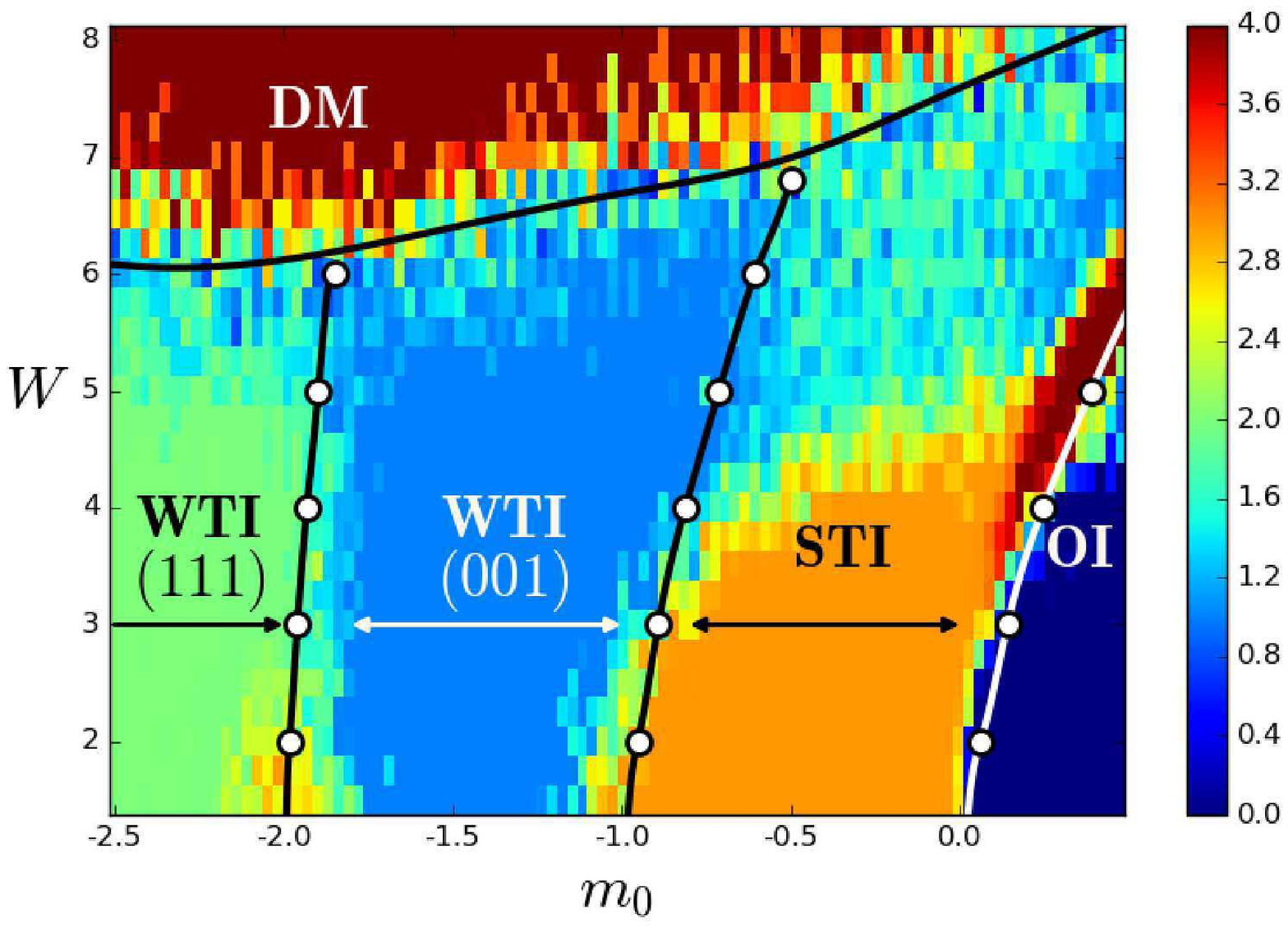}
      \hspace{1.6cm} (a)
     \end{center}
 \end{minipage}
 \begin{minipage}{0.48\hsize}
  \begin{center}
   \includegraphics[angle=0,width=0.9\textwidth]{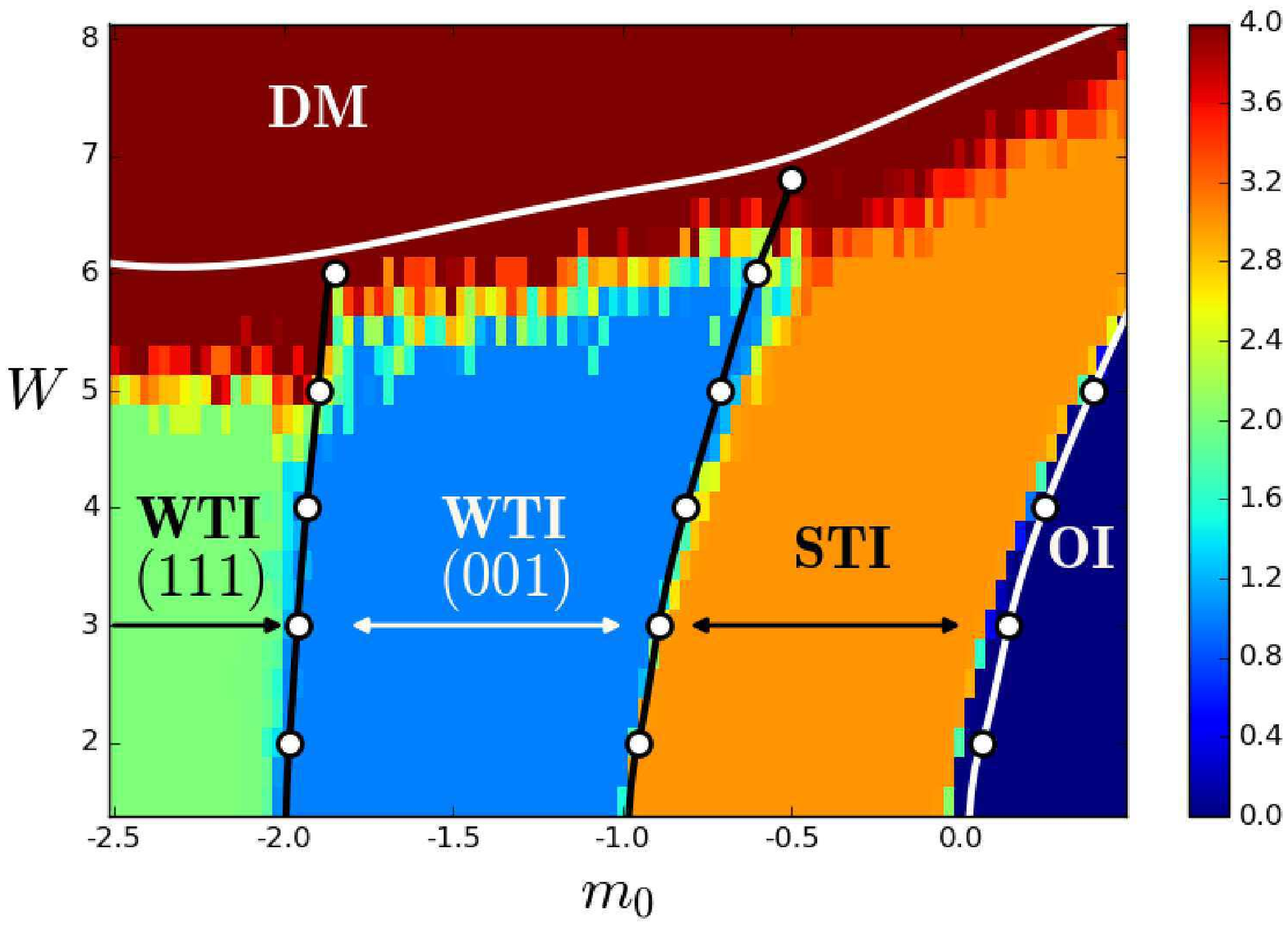}
      \hspace{1.6cm} (b)
     \end{center}
 \end{minipage}
  \end{tabular}

 \caption{(Color online) Color map of $P_\mathrm{OI}, P_\mathrm{W001}, P_\mathrm{W111}, P_\mathrm{STI}$, and $P_\mathrm{DM}$.
  (a) Results of training based on real-space wave functions, and (b) those based on Fourier space wave functions [see Eq. (\ref{eq:Fourier})].
 The intensity $0\times P_\mathrm{OI}+ 1\times P_\mathrm{W001} +2\times  P_\mathrm{W111}+3\times  P_\mathrm{STI}
 +4\times  P_\mathrm{DM}$ is plotted.
 The vertical axis is the strength of disorder $W$, whereas the horizontal axis is the mass parameter $m_0$.
 The shifts of the phase boundaries OI/STI, STI/WTI(001),  and 
 WTI(001)/WTI(111) by randomness are clearly seen.  The arrows indicate the lines along which machine learning for
 STI, WTI(001), and WTI(111) has been performed.
   Solid lines are results from the transfer matrix estimate.\cite{Kobayashi19}
}
\label{fig:phaseDiagramTI}
\end{center}
\end{figure}

To prepare the training data, we set $W=3.0$ and
varied $m_0\in [-2.8,-2]$ to teach the features of WTI(111),
$m_0\in [-1.8,-1]$ to teach the features of WTI(001),
and $m_0\in [-0.8,0]$ to teach those of STI.
These training parameters are along lines in the $W$-$m_0$ plane,
which are shown as  arrows in Fig.~\ref{fig:phaseDiagramTI}(a).
To teach the features of a diffusive metal (DM), we set $W=10.0$ and varied $m_0 \in [-2.5,0.5]$,
whereas for OI, we set $W=3.0$ and varied $m_0 \in [0.2,0.7]$.
Since we do not know the phase diagram for the set of parameters we consider,
we choose the parameters according to the knowledge of the
randomness-free case, $m_{2,z}=0.5$ and $W=0$, together with  information on
the phase diagram for
 the isotropic but disordered case, $m_{2,z}=1$ and $W>0$.\cite{Kobayashi13}
We set $r=0, 1, 2, 3, 4$ [see Eq. (\ref{eq:finalPhase})] to indicate
the labels OI, WTI(001), WTI(111), STI, and DM, respectively.

After teaching 2,000 eigenfunctions in each phase, we  prepared
100$\times$27 eigenfunctions with different $m_0$ (100 values) and $W$ (27 values),
and let the CNN determine which phase each eigenfunction belongs to.
We calculate
the probabilities $P_\mathrm{OI}, P_\mathrm{W001}, P_\mathrm{W111}, P_\mathrm{STI}, $ and
 $P_\mathrm{DM}(=1-P_\mathrm{OI}-P_\mathrm{W001}-P_\mathrm{W111}-P_\mathrm{STI})$
that a given eigenfunction belongs to OI, WTI(001), WTI(111), STI, and DM, respectively.

The probabilities of OI, WTI(001), WTI(111), STI, and DM
are displayed as
a color map in the $W$-$m_0$ plane [Fig.~\ref{fig:phaseDiagramTI}(a)].
We see that the phase boundaries between insulators with different topologies
shift as we increase $W$.
For example, when we start with the OI phase, say $(m_0,W)= (0.3, 1.5)$, and increase the disorder $W$,
we enter into the STI phase at $W\approx 5$.  This is called the topological Anderson insulator (TAI) transition.\cite{Li09a,Groth09,Guo10}
The present method captures the TAI and gives a phase diagram quantitatively consistent with
that obtained by the transfer matrix method.\cite{Kobayashi13,Kobayashi19}
It should be emphasized that training along a few finite 1D lines in a 2D parameter space enables us to draw the
phase diagram.
We also note that the phase boundary between OI and STI is colored red, which indicates
that the phase on the phase boundary is a metal phase.  In fact, the Dirac semimetal continues
to exist on the phase boundary even in the presence of disorder.\cite{Kobayashi14,Syzranov15}

So far, we have considered 3D wave functions in real-space, but with small additional numerical costs,
the wave functions in the Fourier space ($k$-space) are calculated,
\begin{equation}
\label{eq:Fourier}
\psi(k_x,k_y,k_z,i)=\int \mathrm{d}x\, \mathrm{d}y\, \mathrm{d}z\,\, \psi(x,y,z,i)\,
\exp[i (k_y y+k_z z)]\,\sin(k_x x) \,,
\end{equation}
where the fixed boundary condition in the $x$-direction is taken into account, and
$k_y=2\pi n_y/L_y, \, k_z=2\pi n_z/L_z$, and $k_x=\pi n_x/(L_x+1)$, 
with integers $n$ satisfying $0\le n_\mu< L_\mu\, (\mu=y, z)$, and $1\le n_x\le L_x$.
We can also work with the hybrid space, where we Fourier transform the wave functions
only in the $y$- and $z$-directions,
\begin{equation}
\label{eq:Fourier2}
\psi(x,k_y,k_z,i)=\int\, \mathrm{d}y\, \mathrm{d}z\,\, \psi(x,y,z,i)\,
\exp[i (k_y y+k_z z)] \,.
\end{equation}
Now, we can train the CNN by using $|\psi(k_x,k_y,k_z)|^2=\sum_i^4 |\psi(k_x,k_y,k_z,i)|^2$ 
or $|\psi(x,k_y,k_z)|^2=\sum_i^4 |\psi(x,k_y,k_z,i)|^2$ as 3D images, and draw the phase diagram
in exactly the same way as in the case of real-space analyses.
The obtained phase diagram is shown in Fig.~\ref{fig:phaseDiagramTI}(b), where the colors
change more sharply and clearly when the phase changes between insulators with
different topologies.

Before concluding this subsection, we note that
the standard method of using the transfer matrix\cite{Kobayashi13} to determine the phase diagram
in the presence of disorder
breaks down for the choice of parameters $t=m_{2,\mu}$, where $\mu$ is the direction along
the transfer matrix multiplication (see Sect.~\ref{sec:tmm}).
This is because the transfer matrix connecting
a layer to the next layer is not invertible for $t^2-m_{2,\mu}^2=0$.\cite{RyuNomura:3DTI}
The method presented in this subsection, therefore, has wider applicability.\cite{Tomoki17}

\subsubsection{3D Weyl semimetal}
We next consider the 3D Weyl semimetal (WSM)\cite{Murakami07,Wan11}.
One way of realizing the 3D WSM is to consider 2D
Chern insulators (CIs)\cite{Dahlhaus11,Liu16,Chang16}
and stack them in the $z$-direction.\cite{Liu16,Yoshimura16}
We begin with a spinless
two-orbital tight-binding model on a square lattice,
which consists of an $s$-orbital and a $p\equiv p_x+ip_y$ orbital,~\cite{Qi08}
and stack them in the $z$-direction to form a cubic lattice,
\begin{align}
H = & \sum_{{\bm x}} \left(
(\epsilon_s + v_s({\bm x})) c^{\dagger}_{{\bm x},s} c_{{\bm x},s}
+ (\epsilon_p + v_p({\bm x})) c^{\dagger}_{{\bm x},p} c_{{\bm x},p}\right)   \nonumber \\
 +& \sum_{{\bm x}}\Big(-\sum_{\mu=x,y} (
t_s c^{\dagger}_{{\bm x} + {\bm e}_{\mu},s} c_{{\bm x},s}
- t_p c^{\dagger}_{{\bm x} + {\bm e}_{\mu},p} c_{{\bm x},p}) \nonumber \\
+&   t_{sp}
(c^{\dagger}_{{\bm x}+{\bm e}_x,p}
- c^{\dagger}_{{\bm x} - {\bm e}_x,p})  \!\ c_{{\bm x},s} 
-  it_{sp}
(c^{\dagger}_{{\bm x}+{\bm e}_y,p}
- c^{\dagger}_{{\bm x} - {\bm e}_y,p})  \!\ c_{{\bm x},s}
+{\rm H.c.}\Big)  \nonumber \\
-& \sum_{{\bm x}} \Big( t^{\prime}_s c^{\dagger}_{{\bm x} + {\bm e}_{z},s} c_{{\bm x},s}
+ t^{\prime}_p c^{\dagger}_{{\bm x} + {\bm e}_{z},p} c_{{\bm x},p} + {\rm H.c.} \Big)
\,, 
\label{eq:WSMtb1}
\end{align}
where $\epsilon_s$,  $v_s({\bm x})$, $\epsilon_p$, and $v_p({\bm x})$ 
denote the atomic energies and disorder potentials for the $s$- and $p$-orbitals, respectively.
Both $v_s({\bm x})$ and $v_p({\bm x})$ 
are uniformly distributed within $[-W/2,W/2]$ with
an independent probability distribution. $t_s$, $t_p$, and $t_{sp}$ are
 transfer energies between neighboring $s$-orbitals, $p$-orbitals, and that between
$s$- and $p$-orbitals, respectively.
$t^{\prime}_{s}$ and $ t^{\prime}_{p}$ are 
interlayer transfer energies, i.e., hopping elements in the $z$-direction.

In the absence of randomness, the Hamiltonian matrix is expressed in $k$-space as
\begin{equation}
H({\bm k}) = a_0 \sigma_0 + {\bm a}\cdot {\bm \sigma}
\label{eq:wsm1}
\end{equation}
with ${\bm \sigma}=(\sigma_x,\sigma_y,\sigma_z)$  Pauli matrices and
\begin{align}
a_0({\bm k}) &= \frac{\epsilon_s+\epsilon_p}{2} + (t_p-t_s) (\cos k_x + \cos k_y)
- (t^{\prime}_s + t^{\prime}_p) \cos k_z, \nonumber \\
a_3({\bm k}) &=  \frac{\epsilon_s-\epsilon_p}{2} - (t_p+t_s) (\cos k_x + \cos k_y)
- (t^{\prime}_s - t^{\prime}_p) \cos k_z, \nonumber \\
a_1({\bm k}) &= -2t_{sp} \sin k_y, \nonumber \\
a_2({\bm k}) &= -2t_{sp} \sin k_x. \nonumber
\end{align}

As in Ref.\cite{Liu16}, we set $\epsilon_s=-\epsilon_p\,,\, \epsilon_s-\epsilon_p=-2(t_s+t_p)$,
$t^{\prime}_s=-t^{\prime}_p>0$, $t_s=t_p>0$, and $t_{sp}=4t_s/3$,
and take $4t_s$ as the energy unit.
The dimensionless interlayer coupling is defined as
\begin{equation}
\label{eq:WSMbeta}
\beta \equiv \frac{t^{\prime}_p-t^{\prime}_s}{2(t_s+t_p)}\,.
\end{equation}
In the absence of randomness,
this choice of parameters realizes CI with a  band gap in the 2D limit,
$\beta =0$.
As long as $1/2> |\beta|\ge 0$, the energy band remains gapped, and the system
continues to be CI. The system enters into the 3D WSM phase for $|\beta|>1/2$.\cite{Liu16}

In the presence of randomness, four phases appear; CI, WSM, DM, and the Anderson insulator.
Here, we focus on the first three phases by considering $W<2.5$ and $0.3<\beta<0.6$.
(The Anderson insulator phase appears in the larger $W$ region.)
Actually, WSM can be further classified according to the number of Weyl node pairs.
For example, WSM(II) has two pairs of Weyl nodes, $k=\pm k_1\,,\,\pm k_2$, where the energy in a clean system
is $E(\pm k_i)=0 \, (i=1,2)$.
In our set of parameters, we expect two or three pairs of Weyl nodes, so
we define here $P_\mathrm{WSMII}$ and  $P_\mathrm{WSMIII}$ instead of $P_\mathrm{WSM}$ alone.

As in the previous subsection, let us first consider the features of states near $E=0$ in the case of  a small randomness
with periodic boundary conditions imposed in the $y$- and $z$-directions,
whereas the fixed boundary condition is imposed in the $x$-direction.
In the case of a CI,
at $E\approx 0$, edge states run in the $y$-direction [Fig.~\ref{fig:wfATReal}(f)].
On the other hand,  in WSM,
the surface states corresponding to the Fermi arcs\cite{Okugawa14} appear on surfaces normal to the $x$-direction [Figs.~\ref{fig:wfATFourier}(g) and \ref{fig:wfATFourier}(h)].
Owing to the presence of the bulk Weyl node near the same energy $E=0$,
the left and right surface states are coupled,
and the high-amplitude regions appear on both surfaces normal to the $x$--direction
[Figs.~\ref{fig:wfATReal}(g) and \ref{fig:wfATReal}(h)].

\begin{figure}[htb]
  \begin{center}
     \begin{tabular}{cc}
 \begin{minipage}{0.48\hsize}
  \begin{center}
   \includegraphics[angle=0,width=0.9\textwidth]{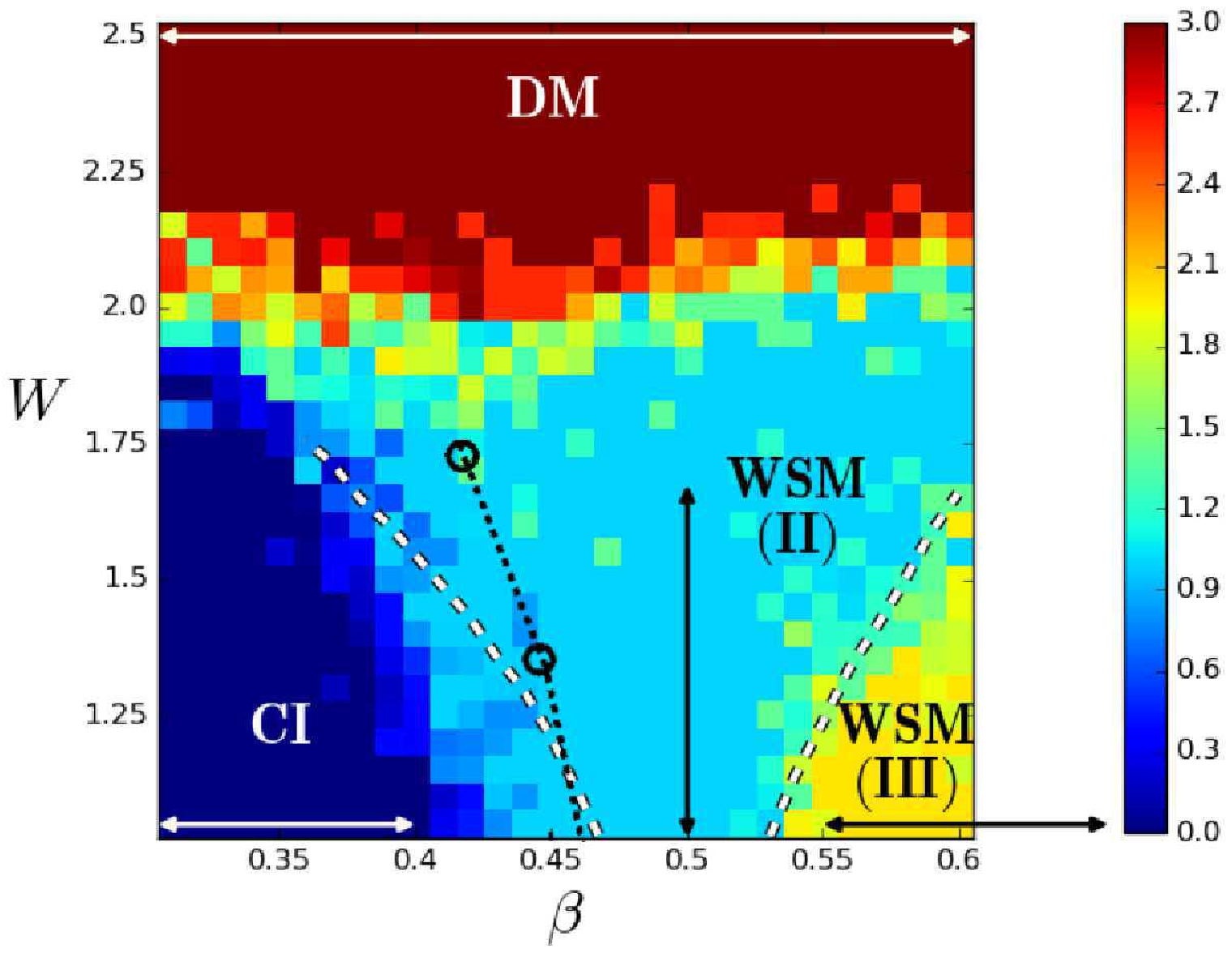}
      \hspace{1.6cm} (a)
     \end{center}
 \end{minipage}
 \begin{minipage}{0.48\hsize}
  \begin{center}
   \includegraphics[angle=0,width=0.9\textwidth]{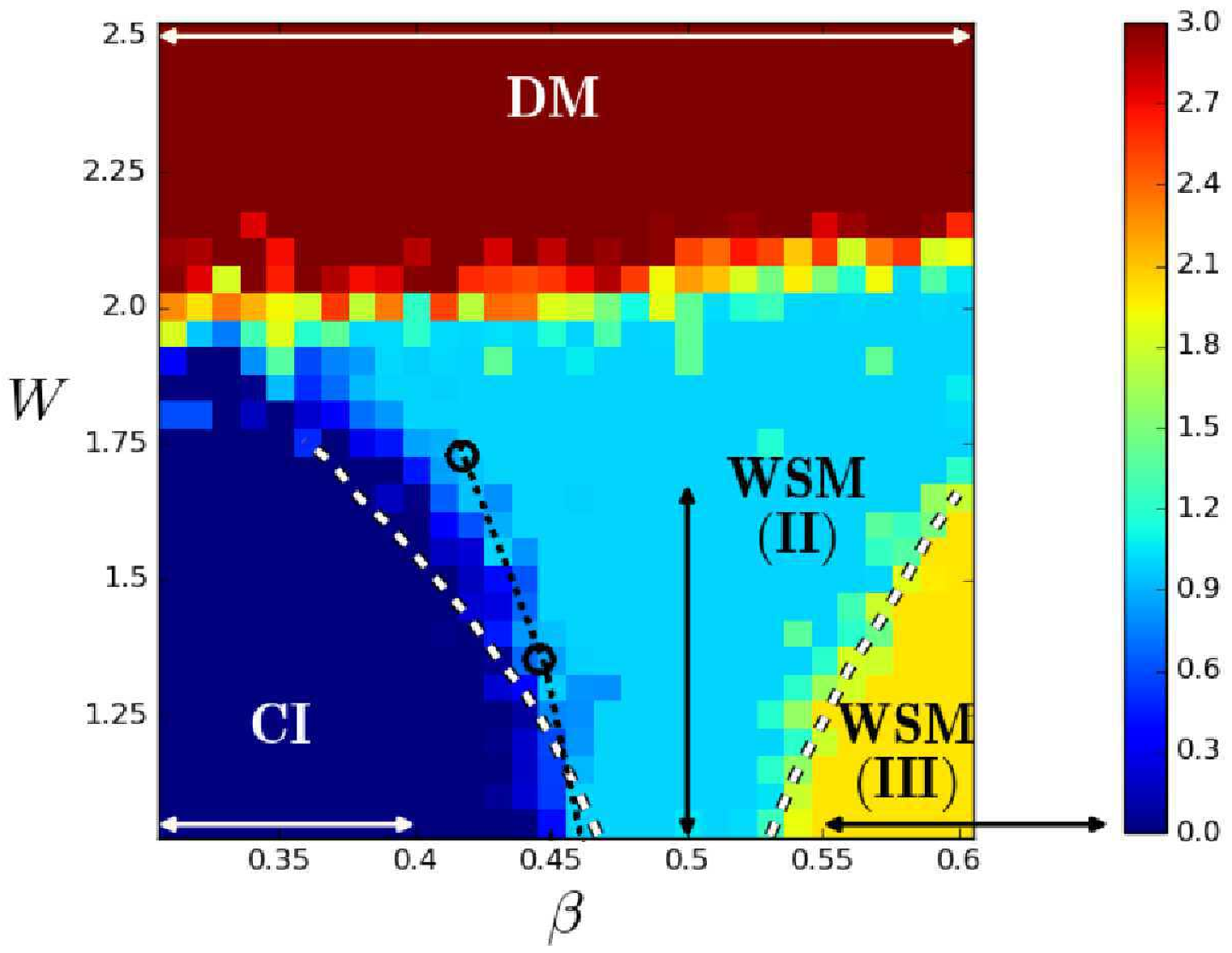}
      \hspace{1.6cm} (b)
     \end{center}
 \end{minipage}
  \end{tabular}

 \caption{(Color online) Color map of $P_\mathrm{CI}, P_\mathrm{WSMII}, P_\mathrm{WSMIII}$, and $P_\mathrm{DM}$.
  (a) Results of training based on real-space wave functions, and (b) those based on Fourier space wave functions.
  The intensity $0\times P_\mathrm{CI}+ 1\times P_\mathrm{WSMII} +2\times  P_\mathrm{WSMIII}+3\times  P_\mathrm{DM}$ is plotted.
 The vertical axis is the disorder strength  $W$, whereas the horizontal axis is the dimensionless interlayer coupling $\beta$.
  The black dotted lines are results from the transfer matrix estimate,\cite{Liu16} whereas the white dashed lines are the estimate by the self-consistent
  Born approximation.\cite{Ominato14,Liu16}
  The arrows indicate the parameters along which the training data have been prepared.
  }
\label{fig:phaseDiagramWSM}
\end{center}
\end{figure}

We set $W=2.5$ and
varied $\beta\in [0.3,0.6]$ to teach the features of DM, and
$W=1.0,\beta\in [0.3,0.4]$ to teach those of CI.  For WSM(II), we set $\beta=0.5, W \in [1.0,1.7]$
and for WSM(III), $W=1.0, \beta\in [0.55,0.65]$.
For each phase, we have diagonalized $32\times 32\times 32$ systems, and prepared 1,000  samples for teaching the features of eigenfunctions.
We then varied $\beta$ and $W$ and let the trained CNN calculate
the probabilities $P_\mathrm{CI}, P_\mathrm{WSMII},P_\mathrm{WSMIII}$, and $P_\mathrm{DM}(=1-P_\mathrm{CI}-P_\mathrm{WSMII}-P_\mathrm{WSMIII})$ that a given eigenfunction belongs to CI, WSM(II), WSM(III), and DM,
respectively.
We then draw  a color map in the $W$-$\beta$ plane, as shown in Fig.~\ref{fig:phaseDiagramWSM},
which quantitatively reproduces the phase diagram obtained by the transfer matrix method.\cite{Liu16}

As in the case of the topological insulator, the phase diagram based on the supervised training of
real-space wave functions [Fig.~\ref{fig:phaseDiagramWSM}(a)] is noisy.
Again, the situation is improved if we work in the $k$-space [Fig.~\ref{fig:phaseDiagramWSM}(b)].

\section{Summary and Concluding Remarks}
In this study, we have shown how the neural network is used to draw various phase diagrams
in the quantum phase transitions.
We have used the wave function as an input and determined the material phase in which
the wave function is obtained.
Both real-space and $k$-space wave functions are used.
Note that numerical diagonalization is carried out in the real-space where the Hamiltonian
becomes sparse, and the $k$-space wave function can be  calculated with small extra numerical costs,
since we focus on the wave functions closest to the band center for topological systems.

In the case of topological insulators, the phase transitions between different topological phases are more clearly
detected by the CNN if we work in the $k$-space.  The phase boundary between the
metal phase and ordinary/topological insulators,
however, does not agree well with the transfer matrix calculation.
The phase boundary between metal and insulators is
more accurate if we work in the real-space.
This is also the case for the Anderson metal--insulator transition, where working in real-space
is better than that in $k$-space.
Whether working in $k$-space is better than that in real-space, therefore, depends
on the nature of the transition.

One of the advantages of this approach is the wider applicability;
it can be applied to the cases
where the transfer matrix method breaks down (see Sect.~\ref{sec:tmm}).  One might think of
switching from the transfer matrix method
to the iterative Green's function method\cite{Ando85} to avoid the inversion
of the matrix connecting one layer to the next [see Eq.~(\ref{eq:tmm})].
In the case of quantum percolation, however, since we consider the bar geometry
where the cross section is finite, the largest cluster in the first layer is often
truncated after a certain number of iterative calculations, which leads to the
breakdown of the Green's function method.
The downside of the present method is that it requires  diagonalizations of Hamiltonians,
so the analyses of higher dimensional systems become more difficult than the
conventional methods.

We have used the supervised training, i.e., the critical point is known for
certain regions of a phase diagram, and the trained neural network is
applied to other regions of the phase diagram where the critical points are unknown.
We have chosen the training regions close to the critical point, but not too close.
This is because training in regions far from the critical point is trivial,
and in regions too close to the critical point, the length scales are
so large that we cannot distinguish one phase from the other
and the labeling of the phase is meaningless.
Another approach to determine the critical point is that we assume
the critical point $x_c$, vary $x_c$ and observe how the training scores
change.\cite{Nieuwenburg17,Li17}
Although the estimate of the critical point is less precise than the conventional
method, the idea may be applied to problems where the conventional method
is difficult to apply.

The accurate estimate of the critical exponent for the quantum phase transition,
 such as the quantum Hall effect transition,\cite{Li17} is still underway, but
is an important problem left for the future.
One might think  that the probability $P$, as in Fig.~\ref{fig:AT}(a), around
the critical point changes more rapidly with the increase in the
system size $L$, with the slope proportional
to $L^{1/\nu}$, $\nu$ being the critical exponent for
the divergence of the length scale.
We, however, do not know how we should
change hyperparameters such as 
 convolution/pooling kernel sizes and the depth of the network as we change $L$.
Take the 3D Anderson transition as an example.
For $L=40$, we used convolution kernel size 5 and pooling kernel size 2, and
the network consists of  6 convolutions, 3 pooling, and 2 fully connected layers (see Table \ref{tb:CNN3Danderson}).
When we simulate larger systems, say, $L=80$, should we use the same
hyperparameters, or should we increase the kernel sizes and network depth?
In the case of the latter, how?
Unless we understand the effect of kernel sizes and network depth on finite size scaling,
a reliable estimate of $\nu$ and its error bar is difficult.

One of the important quantities that is often used in the context of the localization--delocalization problem\cite{Evers00,Mirlin00} is
the inverse participation ratio (IPR),\cite{Weaire77}
\begin{equation}
\label{eq:ipr}
\mathrm{IPR}=\int{\rm d}^d x |\psi (\bm{x})|^4 \,,
\end{equation}
which indicates the inverse of the ``volume" of the wave function.
For example, the IPR of the localized wave function $\psi\sim \exp(-|\bm{x}-\bm{x}_0|/\xi)$ is
size-independent and of the order $1/\xi^d$,
whereas that of the plane wave $\psi\sim \exp(i\bm{k}\cdot\bm{x})/\sqrt{L^d}$ is $1/L^d$.
At the Anderson transition, the size dependence of the IPR is $L^{-d_2}$, where $d_2$ is the fractal dimension
$0<d_2<d$.

\begin{figure}[htb]
  \begin{center}
   \includegraphics[angle=0,width=0.45\textwidth]{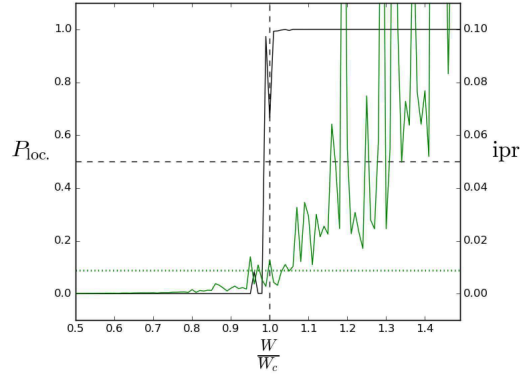}
 \caption{(Color online) Disorder dependence of IPR (green line) and the probability of localized phase (black line) $P_\mathrm{loc}$
 for 3D Anderson model.
 The horizontal axis $W$ indicates the disorder strength.
 The dashed lines indicate $P_\mathrm{loc}=0.5$ and $W=W_c$, whereas
 the green dashed line indicates the average of IPR
 at $W_c$.}
\label{fig:iprVsCnn}
\end{center}
\end{figure}

We diagonalize the
3D Anderson model while changing the strength of disorder $W$, pick up the eigenstate closest to
the band center, $E=0$, and calculate the IPRs and plot them
 in Fig.~\ref{fig:iprVsCnn}.
The same eigenfunction is input to the CNN to calculate the probability of localization $P_\mathrm{loc}$.
As seen in the figure, IPRs are strongly fluctuating, whereas the
CNN outputs are less fluctuating and  change sharply around $W_c$.

\begin{figure}[htb]
  \begin{center}
     \begin{tabular}{cc}
 \begin{minipage}{0.48\hsize}
  \begin{center}
   \includegraphics[angle=0,width=0.9\textwidth]{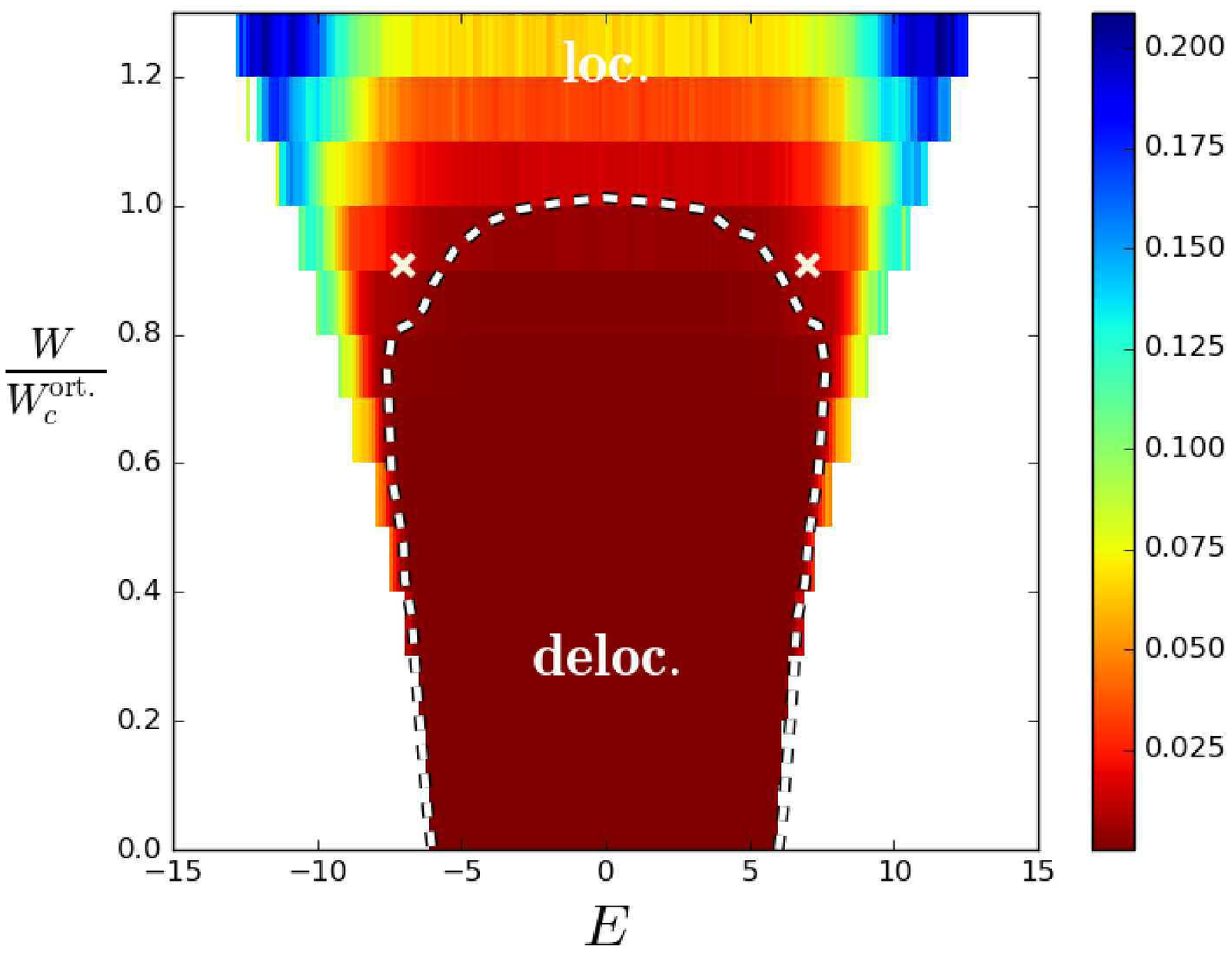}
      \hspace{1.6cm} (a)
     \end{center}
 \end{minipage}
 \begin{minipage}{0.48\hsize}
  \begin{center}
   \includegraphics[angle=0,width=0.9\textwidth]{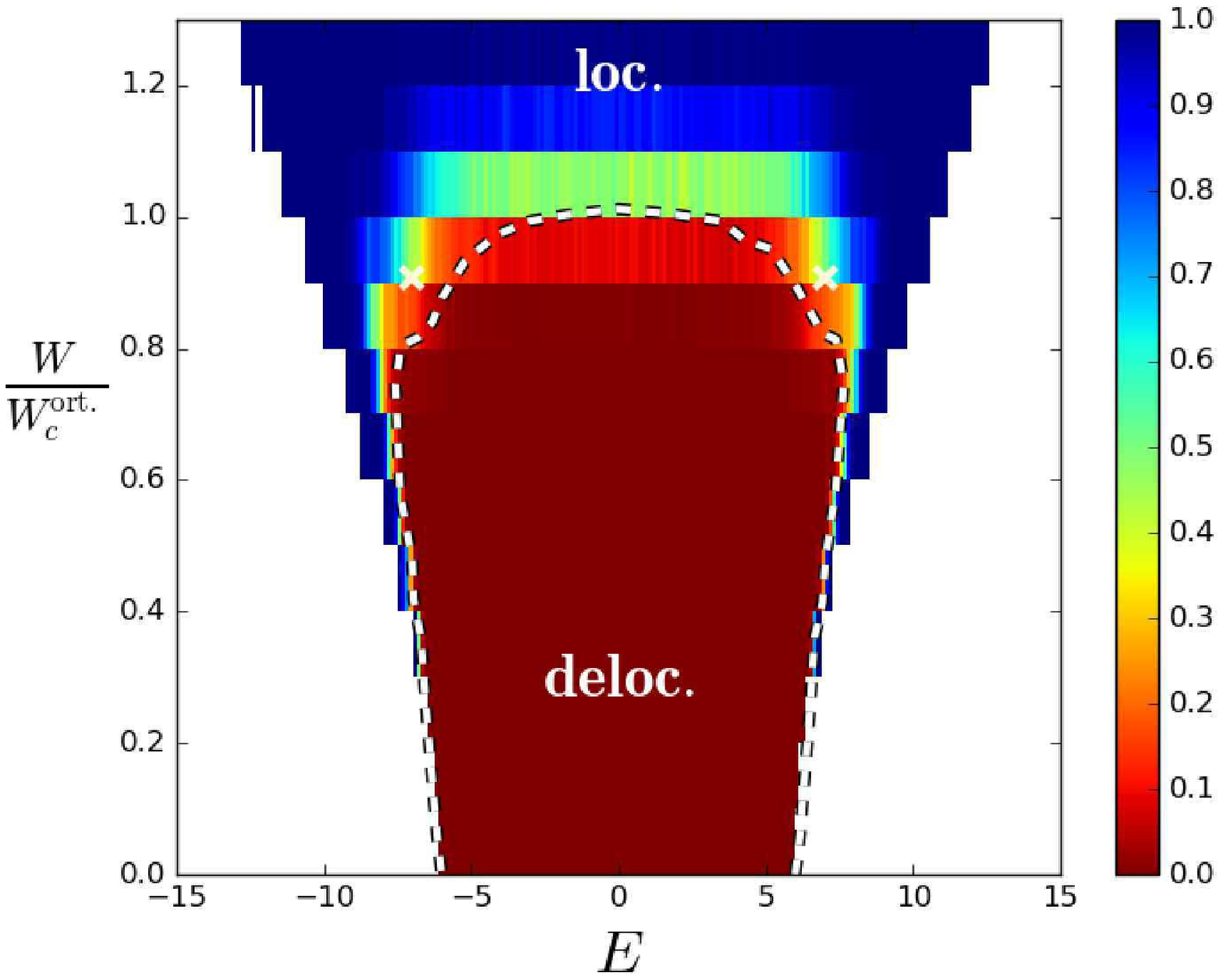}
      \hspace{1.6cm} (b)
     \end{center}
 \end{minipage}
  \end{tabular}

 \caption{(Color online) (a) Similar to Fig.~\ref{fig:AT}, but IPR averaged over five samples and small energy bins are plotted.
(b) Similar to (a) but $\langle \Theta (\mathrm{IPR}_E-\mathrm{IPR}_c)\rangle$ is plotted.}
\label{fig:IPRPhaseDiagram}
\end{center}
\end{figure}

We now take the average of IPR, $\langle \mathrm{IPR}_E\rangle$, where $ \mathrm{IPR}_E$ is the
average of IPR over a small energy bin around the energy $E$, and $\langle \cdots \rangle$ is the sample average.
The results are plotted in Fig.~\ref{fig:IPRPhaseDiagram}(a), where the phase transition
is still difficult to observe.

At $E=0, W=W_c$, we can calculate the critical value of IPR, $\mathrm{IPR}_c=\langle \mathrm{IPR}\rangle_c$.
Using this value, we try to judge that the states are metal if $\mathrm{IPR}_E<\langle \mathrm{IPR}\rangle_c$
and are insulators if $\mathrm{IPR}_E>\langle \mathrm{IPR}\rangle_c$.
In Fig.~\ref{fig:IPRPhaseDiagram}(b), we plot 
$\langle\Theta (\mathrm{IPR}_E-\mathrm{IPR}_c)\rangle$ where the sample average is taken, which resembles Fig.~\ref{fig:AT}, but the phase boundary is not as sharp as in Fig.~\ref{fig:AT}.
Thus, the CNN has some advantage over IPR analysis. The biggest advantage, however,
 is that we do not need to discover IPR to characterize the Anderson localization.

So far, we have concentrated on the static properties of wave functions.
Another quantity that changes its behavior across the transition is the diffusion property,
which is related to the dynamics of wave packets.
In the metal (delocalized) phase, initially localized wave packets begin to be extended with time $t$,
whereas in the insulator (localized) phase, initially localized wave packets remain localized.
This diffusion property is characterized by the time evolution of the diffusion length $r$ of wave packets, which behaves as
\begin{equation}
\langle r^2(t) \rangle \sim
\left \{ \begin{array}{cc}
Dt , & \mathrm{metal},\\
\ t^\alpha, & \mathrm{critical},\\
\xi^2, & \mathrm{insulator},
 \end{array}\right.
\end{equation}
where $D$ is the diffusion constant and $\alpha=2/d$ for the Anderson transition.\cite{Ohtsuki97}  One way to detect such changes in time-dependent behaviors is
to use the CNN for analyzing $r(t)$ vs $t$ images.  Another way is to use a
recurrent neural network, which is
widely used for analyzing time series.\cite{Nieuwenburg18,Bo19}

The deep neural network used here is a tool for classifying the phase
and is regarded as a blackbox.
The properties of the neural network themselves are also interesting to study from the physics view point,
\cite{Obuchi16,Lin17,Luchak17,Barra18,KochJanusz18,Wetzel17b,Li18,Morningstar18,Kaubruegger18,Ortegon18,Ringel18,Wu18,
Kenway18,Hartnett18,Kim18,Puskarov18,Rao18,Greitemann19,Rotskoff18,BaityJesi18,Karakida18,Aubin18,Casert18,Geiger19,Wang19,Nguyen19,Hashimoto19,Mehta19,Araujo19,Ben-David19,Cohen19,Baldassi19,Matsumoto19,Suezen19,Kashiwa19}
especially  in relation to the renormalization group\cite{Beny13,Mehta14,Foreman18,Glasser18,Li18b,Iso18,Funai18,Koch19,Hu19a} 
and tensor network.\cite{Aoki16,Stoudenmire16,Suzen17,You18,Liu18a,Stoudenmire18,Glasser18b,Biamonte18,Liao19,Ran19}
The vulnerability of phase determination against adversarial perturbation is also an interesting
topic.\cite{Szegedy13,Jiang19}
Whether the neural network can learn formula such as the winding number is an interesting question,
and in fact, it seems to be the case.\cite{Zhang18,Sun18}
It is natural to apply machine learning to quantum computers.\cite{Lloyd18,Bukov18,Baireuther18,Torlai18,Uvarov19,Teoh19,Pepper19}
as well to apply quantum algorithms to machine learning.\cite{Biamonte17,Lau17,Ohzeki18,Andreasson18}

\bigskip
\noindent
{\bf Acknowledgements}
The authors would like to thank Dr. Tomoki Ohtsuki for prior collaboration on these topics.
They also thank Dr. Koji Kobayashi, Dr. Ken-Ichiro Imura, Dr. Akinori Tanaka, Dr. Akio Tomiya, Professor Ryuichi Shindou, Professor Koji Hashimoto, Professor Masatoshi Imada, Professor Tohru Kawarabayashi, Professor Victor Kagalovsky, Professor Ferdinand Evers, and Professor Keith Slevin for useful discussions. This work was partly supported by JSPS KAKENHI Grant Nos. JP15H03700, JP17K18763,  16H06345, and 19H00658.

\appendix
\renewcommand{\thetable}{\Alph{section}.\arabic{figure}}
\renewcommand{\thefigure}{\Alph{section}.\arabic{table}}
\def\thesection{Apendix\Alph{section}}
\setcounter{equation}{0} 
\setcounter{figure}{0} 
\setcounter{table}{0} 

\section{}
\subsection{Typical 3D wave functions}
\label{sec:wf}
\begin{figure}[htb]
  \begin{center}
\includegraphics[angle=0,width=0.8\textwidth]{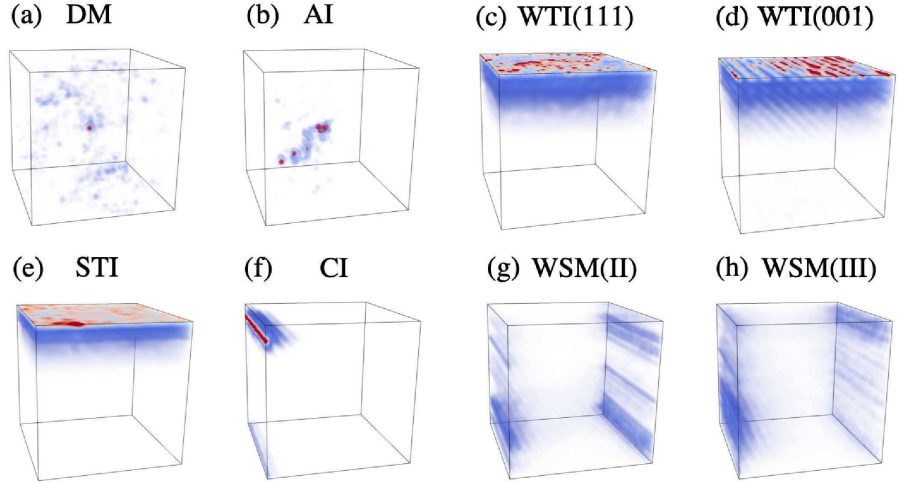}
 \caption{(Color online) 
Typical 3D wave functions in real space in various phases.
(a) and (b) are wave functions in the diffusive metal phase and the Anderson insulator phase, respectively.
 (c)--(e) are for WTI(111),  WTI(001), and STI phases in 3D topological insulators, whereas
  (f)--(h) are for the Chern insulator, WSM(II), and WSM(III), respectively.
Periodic boundary conditions are imposed in all the directions in (a) and (b),
whereas in (c)--(h), fixed boundary condition is imposed in a direction where edge/surface states are observed.
}
\label{fig:wfATReal}
\end{center}
\end{figure}

\begin{figure}[htb]
  \begin{center}
\includegraphics[angle=0,width=0.8\textwidth]{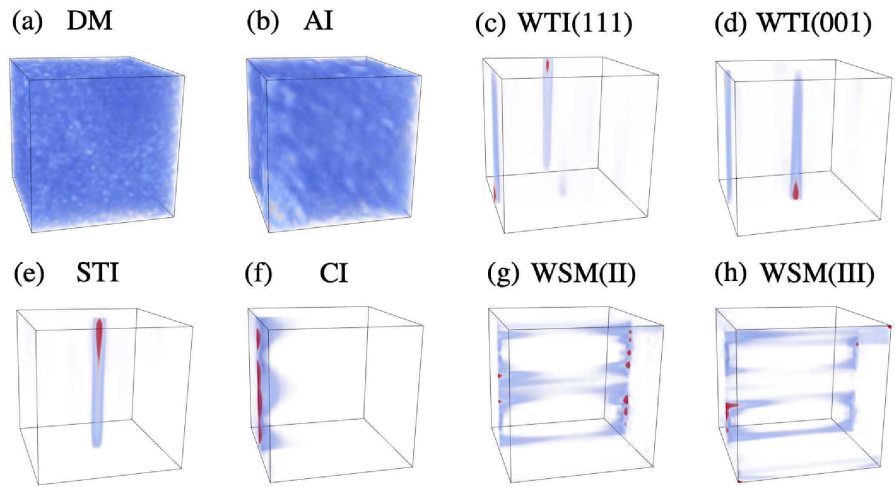}
 \caption{(Color online) 
As in Fig.~\ref{fig:wfATReal}, but in Fourier space.
Equation~(\ref{eq:Fourier}) is used for (c)--(e), and Eq.~(\ref{eq:Fourier2})
is used for (f)--(h).  Fermi arcs are observed in (g) and (h).}
\label{fig:wfATFourier}
\end{center}
\end{figure}

We show below the typical 3D wave functions for a diffusive metal, Anderson insulator, weak and
strong topological insulators as well as Weyl semimetals (Figs. \ref{fig:wfATReal} and \ref{fig:wfATFourier}).
We observe that for the case of the Anderson transition, distinguishing the metal wave function (a) from the insulating one (b)
is easier in real space, whereas in the topological systems (c)--(h), Fourier-transformed wave functions are more informative
in the sense that we can distinguish surface states of different topological phases more easily.

\subsection{CNN hyperparameters}
\label{sec:hyperparameters}
As mentioned in Sect.~\ref{sec:method}, the CNN has parameters that are
not optimized  during the course of supervised training.
These parameters, so-called hyperparameters, must be selected in advance so that the CNN determines the quantum phases for the
validation set with high accuracy.
Our selections of the parameters for Sect.~\ref{sec:results} are shown Tables \ref{tb:CNN3Danderson} to \ref{tb:CNNtopo}.
In the tables, the ``kernel size'' corresponds to the size of the cell cut out from the images
in the previous layer.
When the ``padding'' is True, zero padding, namely, adding zeroes to the peripherals of the input,
is applied so that the output shape is the same as the input shape,
whereas the output shape decreases through the convolution layer when the ``padding'' is False.
To be more specific, when the kernel size is $m$ with the input linear dimension $L$, the output is $L-m+1$ for padding False,
whereas the output size remains $L$ for padding True.
To avoid overfitting, the dropout process, which randomly drops half of the inputs, has been implemented
after each pooling layer as well as the fully connected layer, except for the last layer.
For the 4D CNN, the Adam method was used to minimize the cross entropy, 
whereas the AdaDelta method was used for the others.
\begin{table}[htb]
 \caption{
 Hyperparameters of the CNN used for 3D Anderson model and quantum percolation (Sect.~\ref{sec:results}.1).
 No bias parameters are included in the weight parameters.
 }
  \begin{tabular}{lccccc} \hline  \hline
layer class & channel number & kernel size & padding & stride value & output shape \\ \hline
input & & & & &$(1, 40, 40, 40)$ \\
convolutional-1 &   64 & $5\times5\times5$ & False & 1 & $(64, 36, 36, 36)$  \\
convolutional-2 &   64 & $5\times5\times5$ & True   & 1 & $(64, 36, 36, 36)$  \\
pooling-1        &    -   & $2\times2\times2$ &    -      & 2 & $(64, 18, 18, 18)$  \\
convolutional-3 &   96 & $3\times3\times3$ & False & 1 & $(96, 16, 16, 16)$  \\
convolutional-4 &   96 & $3\times3\times3$ & True   & 1 & $(96, 16, 16, 16)$  \\
pooling-2        &    -  & $2\times2\times2$ &     -     & 2 & $(96,   8,   8,   8)$  \\
convolutional-5 & 128 & $3\times3\times3$ & False & 1 & $(128,  6, 6, 6)$  \\
convolutional-6 & 128 & $3\times3\times3$ & True   & 1 & $(128, 6, 6, 6)$  \\
pooling-3        &    -   & $2\times2\times2$ &    -      & 2 & $(128, 3, 3, 3)$  \\
fully connected-1 & & & & &$(1024)$  \\
fully connected-2 & & & & &$(2)$ \\  \hline \hline
\end{tabular}
\label{tb:CNN3Danderson}
\end{table}

\begin{table}[htb]
 \caption{Hyperparameters for 2D CNN used for 2D SU(2) model (Sect.~\ref{sec:results}.3).}
  \begin{tabular}{lccccc} \hline \hline
layer class & channel number & kernel size & padding & stride value & output shape \\ \hline
input & & & & &$(1, 40, 40)$ \\
convolutional-1 &   16 & $5\times5$ & False & 1 & $(16, 36, 36)$  \\
convolutional-2 &   16 & $5\times5$ & True   & 1 & $(16, 36, 36)$  \\
pooling-1        &    -   & $2\times2$ &   -       & 2 & $(16, 18, 18)$  \\
convolutional-3 &   32 & $3\times3$ & False & 1 & $(32, 16, 16)$  \\
convolutional-4 &   32 & $3\times3$ & True   & 1 & $(32, 16, 16)$  \\
pooling-2        &    -  & $2\times2$ &     -     & 2 & $(32,   8,   8)$  \\
convolutional-5 &   64 & $3\times3$ & False & 1 & $(64, 6, 6)$  \\
convolutional-6 &   64 & $3\times3$ & True   & 1 & $(64, 6, 6)$  \\
pooling-3        &    -   & $2\times2$ &      -    & 2 & $(64, 3, 3)$  \\
fully connected-1 & & & & &$(512)$  \\
fully connected-2 & & & & &$(2)$ \\ \hline \hline
\end{tabular}
\label{tb:CNN2D}
\end{table}

\begin{table}[htb]
 \caption{Hyperparameters for 4D CNN used for 4D Anderson model (Sect.~\ref{sec:results}.4).}
  \begin{tabular}{lccccc} \hline \hline
layer class & channel number & kernel size & padding & stride value & output shape \\ \hline
input & & & & &$(1, 16, 16, 16, 16)$ \\
convolutional-1 &   16 & $5\times5\times5\times5$ & True & 1 & $(16, 16, 16, 16, 16)$  \\
pooling-1        &    -   & $2\times2\times2\times2$ &    -      & 2 & $(16, 8, 8, 8, 8)$  \\
convolutional-2 &   16 & $3\times3\times3\times3$ & True & 1 & $(16, 8, 8, 8, 8)$  \\
pooling-2        &    -  & $2\times2\times2\times2$ &     -     & 2 & $(16, 4, 4, 4, 4)$  \\
fully connected-1 & & & & &$(256)$  \\
fully connected-2 & & & & &$(2)$\\ \hline \hline
\end{tabular}
\label{tb:CNN4D}
\end{table}

\begin{table}[htb]
 \caption{Hyperparameters for
 3D CNN for topological materials (Sect.~\ref{sec:results}.5). $L$ is the system size, and $n$  the number of material phases; $L=24,n=5$ for a
 topological insulator, whereas for the Weyl semimetal, $L=32, n=4$.}
  \begin{tabular}{lccccc} \hline  \hline
layer class & channel number & kernel size & padding & stride value & output shape \\ \hline
input & & & & &$(1, L, L, L)$\\
convolutional-1 &   16 & $5\times5\times5$ & True & 1 & $(16, L, L, L)$  \\
convolutional-2 &   16 & $5\times5\times5$ & True   & 1 & $(16, L, L, L)$  \\
pooling-1        &    -   & $2\times2\times2$ &    -      & 2 & $(16, \frac{L}{2},\frac{L}{2},\frac{L}{2})$  \\
convolutional-3 &   32 & $3\times3\times3$ & True & 1 & $(32, \frac{L}{2},\frac{L}{2},\frac{L}{2})$  \\
convolutional-4 &   32 & $3\times3\times3$ & True   & 1 & $(32, \frac{L}{2},\frac{L}{2},\frac{L}{2})$  \\
pooling-2        &    -  & $2\times2\times2$ &     -     & 2 & $(32,   \frac{L}{4},\frac{L}{4},\frac{L}{4})$  \\
fully connected-1 & & & & &$(1024)$  \\
fully connected-2 & & & & &$(n)$ \\  \hline \hline
\end{tabular}
\label{tb:CNNtopo}
\end{table}

\subsection{Breakdown of transfer matrix method}
\label{sec:tmm}
In this subsection, we explain why the transfer matrix method, which is widely used in the study of
Anderson localization,\cite{MacKinnon81,Pichard81,Kramer93,Slevin14}
breaks down in certain lattices such as quantum percolation, fractal lattices,\cite{Asada06}
and topological insulators\cite{RyuNomura:3DTI} as well as the Weyl semimetal.\cite{Liu16}

For simplicity, as in the main text, we consider the Hamiltonian where only the nearest-neighbor couplings are allowed,
and consider a long bar in the $x$-direction with cross section $L_y\times L_z$.  We denote the values of the wave function on the
$n$th cross section normal to the $x$-direction as the $M$-dimensional vector $\Psi_n$,
where $M$ is the degree of freedom on the cross section ($L_y\times L_z\times$ internal degrees of freedom such as spin and orbital).
From the Schr\"odinger equation, we relate $\Psi_{n+1}$ to $\Psi_n$ and $\Psi_{n-1}$ as
\begin{equation}
E \Psi_n=H_n \Psi_n + V_{n,n+1} \Psi_{n+1}+V_{n,n-1} \Psi_{n-1}\,,
\end{equation}
which is rewritten as
\begin{equation}
\label{eq:tmm}
\left(
\begin{array}{c}
 \Psi_{n+1}      \\
 V_{n+1,n}   \Psi_{n}      
\end{array}
\right)
=\left(
\begin{array}{cc}
  V_{n,n+1}^{-1}& 0_M    \\
0_M                        & V_{n+1,n}  
\end{array}
\right)
\left(
\begin{array}{cc}
  E\,I_M-H_n& -I_M    \\
  I_M        & 0_M
\end{array}
\right)
\left(
\begin{array}{c}
 \Psi_{n}      \\
 V_{n,n-1}   \Psi_{n-1}      
\end{array}
\right)\,,
\end{equation}
where $H_n$ is the $M\times M$ Hamiltonian matrix on the $n$th cross section,
and $I_M$ and $0_M$ are unit and zero matrices of dimension $M$, respectively.
The transfer matrix, 
\begin{equation}
T_n=
\left(\begin{array}{cc}
  V_{n,n+1}^{-1}& 0_M    \\
0_M                        & V_{n+1,n}  
\end{array}
\right)
\left(
\begin{array}{cc}
  E\,I_M-H_n& -I_M    \\
  I_M        & 0_M
\end{array}
\right)\,,
\end{equation}
 therefore requires the existence of the inverse matrix, $V_{n,n+1}^{-1}$.

In the quantum percolation, $V_{n,n+1}$ is a diagonal matrix, the elements of which are zero 
when the nearest-neighboring sites in the $x$--direction are disconnected, leading to $\det(V_{n,n+1})=0$.
In the case of a fractal lattice, $V_{n,n+1}$ can be nonsquare matrix.
In the case of topological insulators,  $\det(V_{n,n+1})=((t^2-m_{2,x}^2)/4)^{2L_y L_z}$, so even in the case of a simple cubic lattice,
the transfer matrix method does not apply for the choice of parameters, $t=m_{2,x}$.
Similarly, in our model of the 3D Weyl semimetal, Eq.~(\ref{eq:WSMtb1}), the transfer matrix method breaks down
in the $x$- and $y$-directions when $t_{sp}^2-t_s t_p=0$.

\newpage

\bibliography{ohtsuki19}

\end{document}